# An application of tomographic PIV to investigate the spray-induced turbulence in a direct-injection engine

H. Hill[1], C.-P. Ding[2], E. Baum[2], B. Böhm[2], A. Dreizler[2], B. Peterson[1*]

[1]School of Engineering, Institute for Multiscale Thermofluids, University of Edinburgh, The King's Buildings, Mayfield Road, Edinburgh, EH9 3FD, UK

[2]Fachgebiet Reaktive Strömungen und Messtechnik (RSM), Technische Universität Darmstadt, Otto-Berndt-Straße 3, 64287 Darmstadt, Germany

**\*Corresponding author:** Brian Peterson, The King's Buildings, Mayfield Road, Edinburgh, EH9 3BF, Scotland, UK, Fax: +44 (0)131 650 6554, E-mail: brian.peterson@ed.ac.uk

## Abstract

Fuel sprays produce high-velocity, jet-like flows that impart turbulence onto the ambient flow field. This spray-induced turbulence augments rapid fuel-air mixing, which has a primary role in controlling pollutant formation and cyclic variability in direct-injection engines. This paper presents tomographic particle image velocimetry (TPIV) measurements to analyse the 3D spray-induced turbulence during the intake stroke of a direct-injection spark-ignition (DISI) engine. The spray produces a strong spray-induced jet (SIJ) in the far field, which travels through the cylinder and imparts turbulence onto the surrounding flow. Planar high-speed PIV measurements at 4.8 kHz are combined with TPIV at 3.3 Hz to evaluate spray particle distributions and validate TPIV measurements in the particle laden flow. A comprehensive uncertainty analysis is performed to assess the uncertainty associated with individual vorticity and strain rate components.

TPIV analyses quantify the spatial domain of the turbulence in relation to the SIJ and describe how turbulent flow features such as turbulent kinetic energy (TKE), strain rate (S) and vorticity (Ω) evolve into the surrounding flow field. Access to the full S and Ω tensors facilitate the evaluation of turbulence for individual spray events. TPIV images reveal the presence of strong shear layers (visualized by high S magnitudes) and pockets of elevated vorticity along the immediate boundary of the SIJ. S and Ω values are extracted from spatial domains extending in 1mm increments from the SIJ. Turbulence levels are greatest within the 0-1mm region from the SIJ boarder and dissipate with radial distance. Individual strain rate and vorticity components are analyzed in detail to describe the relationship between local strain rates and 3D vortical structures produced within strong shear layers of the SIJ. Analyses are intended to understand the flow features responsible for rapid fuel-air mixing and provide valuable data for the development of numerical models.

## 1. Introduction

With concerns of anthropogenic climate change, yet increasing demand of passenger vehicles, engineers and scientists are vigorously striving to provide clean, energy efficient (low $CO_2$) vehicles. While electric propulsion vehicles provide opportunities for cleaner tailpipe emissions, it should equally be recognized that improving efficiency from internal combustion (IC) engines is a proven and effective methodology for vehicle $CO_2$ reduction (Alaguamalai 2014). This aspect is important in the overall development of cleaner powertrains, including hybrid technology.

Improved thermal efficiency can be achieved using direct-injection (DI) strategies for spark-ignition (SI) engines. DI strategies offer reduced pumping losses by reduced throttling while the engine load is controlled by high-pressure fuel injection directly into the cylinder (Zhao et al. 1999). Proper fuel mixing is critical to obtain reliable ignition and reduced emissions. Even for *early injection* when fuel is injected during the intake stroke, mixing can be insufficient such that mixtures are still heterogeneous at the end of compression (Snyder et al. 2011). Variations of such mixture distributions cause difficulties in consistent engine performance and emission control (Alger et al. 2004).

The in-cylinder turbulent flow plays a pivotal role in mixture preparation within DI engines. Injection of liquid fuel in excess of 15 MPa imposes spray-induced turbulence that presents strong mixing layers, large velocity gradients, and locally coherent vortical flows that all regulate the success of rapid fuel-air mixing and transport. This spray-induced turbulence can also enhance/disturb the pre-existing flow field, which will positively/negatively affect the mixing process (Stiehl et al. 2013, Peterson et al. 2014, Zeng et al. 2016). There is an evident need to understand the spray-induced turbulent flow to optimize mixture preparation for DI technologies.

Within the spray and engine community there are significant efforts to study the spray-induced flow physics. The majority of fundamental investigations are performed under quiescent flow conditions, which provide well-controlled and simplified boundary conditions for sprays. Under these conditions, researchers have studied spray-induced turbulence near the injector nozzle (termed near field) to study primary breakup mechanisms (Som and Aggarwal 2010, Movaghar et al. 2017). Downstream of the nozzle, in the far field, spray-induced turbulence involves the momentum exchange between droplet and gas phases. Zhang et al. (2014) has described this momentum exchange in quiescent gas environments. Banjaree and Rutland (2015) performed numerical simulations in the far field that resolved the evolution of spray-induced coherent vortical structures from a single nozzle fuel jet. In the far field, these turbulent coherent vortical flow structures surrounding the spray play a primary role in fuel-air mixing.

In engines, the spray consists of multiple fuel jets and the gas flow is not quiescent; the spray-induced flow physics in engines is more complex and must be studied. Within engines, the majority of spray-induced flow studies have focused on spray-flow interactions that describe fuel delivery (Stiehl et al. 2013), mixture and thermal transport (Peterson and Sick 2009, Peterson et al. 2015a), spark-ignition (Dahms et al. 2009, Peterson and Sick 2010), and combustion stability (Peterson et al. 2014, Zeng et al. 2015, Zeng et al. 2016). Flow turbulence produced by sprays is often characterized by Reynolds decomposition methods from which the fluctuating velocity components are considered to represent turbulence (e.g. Aleiferis and Behringer 2017, Zhuang et al. 2017, Clark et al. 2018). While such methods have characterized turbulence with engine performance parameters, the flow decomposition does not adequately describe turbulent flow mechanisms responsible for rapid mixing and transport.

A common limitation of most experimental studies of engine turbulence is the inability to measure 3D velocity gradients, which resolves the full strain rate and vorticity tensors. Tomographic PIV (TPIV) has recently been applied within IC engines to resolve the three-dimensional, three component (3D3C) flow fields, providing access to 3D velocity gradients (Baum et al. 2013, van Overbrüggen et al. 2015). Zentgraf et al. (2016) demonstrated the unique ability to study the instantaneous flow turbulence in engines using TPIV by accessing strain rate and vorticity tensors. Moreover, Peterson et al. (2017) recently demonstrated the application of TPIV to study spray-induced turbulence after injection when droplet distributions were suitable for accurate particle reconstruction. Those measurements quantified the statistical evolution of spray-induced turbulence during mid-compression.

In this paper, TPIV is performed to provide a detailed study of the spray-induced turbulence during the intake stroke of a spray-guided (SG) DISI engine. During intake, the interaction of the spray with its surroundings produces a strong spray-induced jet in the far field of the central symmetry plane. This high-velocity jet, infused with air and droplets, travels downwards through the imaging volume and imparts turbulence onto the surrounding in-cylinder flow. Planar high-speed PIV measurements at 4.8 kHz are combined with TPIV at 3.3 Hz to provide a 2D time-history of the spray-flow preceding phase-locked TPIV measurements. High-speed PIV measurements are also used to assess TPIV measurements. A comprehensive uncertainty analysis is performed to assess the uncertainty associated with vorticity and strain rate components. TPIV analysis focuses on quantifying the spatial domain of the turbulence in relation to the spray-induced jet and describes how turbulent flow features such as strain rate and vorticity evolve into the surrounding flow with time. Individual strain rate and vorticity components are analysed to describe the relationship between local strain rate and 3D vortical structures produced within strong shear layers of the spray-induced jet. Such analyses are intended to understand the flow features that are responsible in the mixing process and provide valuable validation data to develop predictive numerical engine simulations.

## 2. Experimental Setup

Velocimetry measurements were performed in a 4-stroke single-cylinder SG-DISI optical engine operating at 800 RPM. The engine is equipped with a 4-valve pentroof cylinder head, centrally-mounted injector, and centrally-mounted spark plug. Optical access is granted through the quartz-glass cylinder and flat piston. Further details of the engine are described in Baum et al. (2014) and Freudenhammer et al. (2015).

Operating parameters, shown in Table 1, were chosen to mimic a low-load engine operation, but the engine was not operated in the fired mode. Operating conditions were chosen to agree with the comprehensive velocimetry databases for the motored flow (Baum et al. 2014; Freudenhammer et al. 2015; Zentgraf et al. 2016), spray-induced flow (Peterson et al. 2015a; Peterson et al. 2017) and reacting flow (Peterson et al. 2015b; He et al. 2017; Peterson et al. 2019; Ding et al. 2019) associated with this engine. Silicone oil droplets (0.5 μm diameter) were seeded into the intake air for PIV by means of an aerosol generator (AGF 10.0, Palas). However, fuel droplets can also influence velocimetry measurements (see Sect. 3.2). Isooctane was injected through a centrally-mounted, outwards opening piezo-actuated injector (105º spray angle) with 18 MPa injection pressure. The injector operated with 500 μs injection duration and an end-of-injection (EOI) of 277º crank-angle degrees (ºCA) before top-dead-center (bTDC). The amount of fuel injected was 3.6 mg/cycle. This injection event mimics a single-injection typically utilized amongst a multi-injection strategy to avoid wall-wetting.

*Table 1: Engine operating conditions.*

| | |
|---|---|
| Engine speed | 800 RPM |
| Bore, Stroke | 86 mm, 86 mm |
| Compression ratio | 8.7 |
| Fuel ($C_8H_{18}$), EOI | 3.6 mg/cycle / 277º bTDC |
| Inj. Press. / Temp. | 18 MPa / 333 K |
| Intake Press. / Temp. | 95 kPa / 295 K |
| Charge density at EOI | 1.1 kg/m$^3$ |
| HS-PIV image timing (4.8 kHz) | 285º – 271º bTDC |
| TPIV image timing (3.3 Hz) | 274º, 273º, 272º, 271º, 270º bTDC |

Figure 1 shows the experimental setup for the combined TPIV and planar HS-PIV. A dual-cavity frequency-doubled Nd:YAG laser (PIV 400, Spectra Physics, 350 mJ/pulse) operating at 3.3 Hz was used for TPIV. The laser beam passed through a half-wave plate (p-polarized) and two cylindrical lenses to expand and collimate laser light to specify the laser sheet thickness of 5 mm. The light passed through a polarizing beam splitter and another set of cylindrical lenses to expand and collimate the beam to specify the laser sheet width. The laser light was reflected off a 45º mirror in the crankcase to provide a vertically illuminated volume in the engine. Four interline transfer sCMOS cameras (LaVision, Imager sCMOS), with identical 100 mm lenses (Tokina) in Scheimpflug arrangement were arranged circularly around the engine. TPIV camera angles were chosen to provide the maximum range of camera angles suitable for the field-of-view (FOV). The large camera angles between cameras 3 and 4 accommodated the HS-PIV camera placed in between them. Each TPIV camera projection provided an independent line-of-sight information of the illuminated volume (50 x 40 x 4 mm$^3$) centered at z = 0 mm (i.e. cylinder axis). Table 2 provides further information of the detection systems.

*Table 2: Detection system parameters.*

| | |
|---|---|
| TPIV (100 mm lens, x4) | |
| CMOS sensor size | 2560x2160 pixels |
| Pixel size | 6.5 μm |
| Lens aperture setting | *f*# = 16 |
| Magnification, scale | M = 0.28, 0.023 mm/pixel |
| Depth of field | 11.5 mm |
| HS-PIV (100 mm lens) | |
| CMOS sensor size (active) | 1280x800 pixels (864x720 pixels) |
| Pixel size | 20 μm |
| Lens aperture setting | *f*# = 8 |
| Magnification, scale | M = 0.21, 0.094 mm/pixel |
| Depth of field | 4.4 mm |

A second dual-cavity, frequency-doubled Nd:YAG laser (Edgewave, INNOSLAB IS4 II DE, 8 mJ/pulse) operating at 4.8 kHz was used for planar HS-PIV. The laser beam passed through a quarter-wave plate and through a set of focusing optics before being combined with the TPIV laser at the polarizing beam splitter. Only the s-polarized light of the HS-PIV laser was reflected and used for experiments (i.e. 50% of laser energy, 4 mJ/pulse). After the polarizer, the HS-PIV laser light passed through the same focusing optics as the TPIV system. The HS-PIV laser sheet of 1 mm thickness was positioned within the center of the TPIV volume (z = 0 mm position). A CMOS camera (Phantom v711) equipped with a 100 mm lens (Tokina) was placed between TPIV cameras 3 and 4 and imaged onto a 55 x $H$ x 1 mm$^3$ FOV, where $H$ is determined by the piston position.

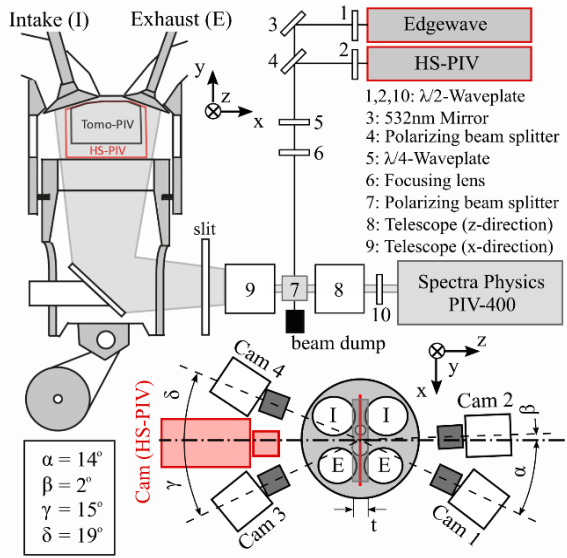

**Figure 1:** Experimental setup of combined HS-PIV / TPIV in the optical engine

All camera and laser systems were synchronized to the engine at 800 RPM. HS-PIV images were recorded at ºCA resolution from 285º bTDC until the ºCA before TPIV images were acquired, providing the 2D2C flow field evolution and droplet distribution before TPIV. This was performed for 300 TPIV images acquired at 270º bTDC. HS-PIV images were not acquired after TPIV because of the sCMOS camera's long exposure time (20 ms); any light from the HS-PIV laser within the second TPIV exposure would negatively bias TPIV measurements. The laser pulse separation ($\Delta t$) for both the HS-PIV and TPIV laser systems was 10 μs to resolve the spray-induced flow and high-velocity intake flow. HS-PIV images were acquired for 288 consecutive cycles, while TPIV images were recorded every 2$^{nd}$ cycle to acquire 300 phase-locked images at 270º bTDC. Limited disk space of the HS-PIV camera (8 GB) prevented the camera from recording more than 288 cycles. This limited the number of synchronized HS-PIV / TPIV sequences to 144 cycles per experiment.

Additional phase-locked TPIV images were taken from 274º-270º bTDC (100 cycles for each ºCA). HS-PIV was not performed for this sequence. These TPIV images were recorded to study the 3D3C spray-induced flow evolution after EOI. Particle distributions were too dense to utilize TPIV before 274º bTDC.

TPIV and HS-PIV were processed with DaVis 8.2.1 (LaVision). Images of a spatially defined 3D target within the engine were used to calibrate images and match viewing planes of each camera system. A 15 pixel sliding minimum subtraction and local intensity normalization were applied during TPIV pre-processing. A volume self-calibration (Wieneke 2008) was performed for 100 images without injection. This provided a remaining pixel disparity of less than 0.2 pixels. 3D particle reconstruction was performed using an iterative Multiplication Algebraic Reconstruction Technique algorithm (FastMART) (Michaelis et al. 2010; Novara et al. 2010). TPIV was calculated by direct volume correlation with decreasing volume size (final size: 64 x 64 x 64 pixels) with 75% overlap, providing a 1.5 x 1.5 x 1.5 mm$^3$ spatial resolution (based on the final interrogation window size) and 0.375 mm vector spacing in each direction. HS-PIV images were cross-correlated with a decreasing window size, multi-pass iterations from 64 x 64 to 32 x 32 pixels with 75% overlap, providing a 3.0 x 3.0 x 1.0 mm$^2$ spatial resolution and 0.75 mm vector spacing in the *x-y* direction. A 3x3 Gaussian smoothing filter was applied to remove noise at spatial scales near the resolution limits for both PIV and TPIV (Fajardo and Sick 2009).

## 3. Velocimetry assessment
### 3.1 Velocimetry with fuel injection

Injection of fuel into the cylinder results in a rapid influx of additional tracers (fuel droplets) into the flow medium which can create a challenging environment for velocimetry applications. This section presents the evolution of the injection event to describe the fuel droplet distribution in relation to the HS-PIV and TPIV timing. This section also presents particle per pixel (ppp) distributions from TPIV images and discusses the implications of ppp with respect to tomographic reconstruction.

Figure 2 presents an instantaneous Mie scattering sequence at selected °CAs to describe a typical droplet distribution following injection. Beneath each Mie scattering image, a corresponding 2D PDF is shown to describe the probability of identifying liquid fuel droplets based on 288 engine cycles. The 2D PDFs are constructed from binary images that spatially identify fuel droplets within the FOV. Fuel droplets were identified by scattering intensities exceeding a threshold of 1000 counts on the 12-bit HS-PIV camera. This threshold value was determined from analysis of the raw Mie images which compared scattering intensities for injection operation against non-injection operation.

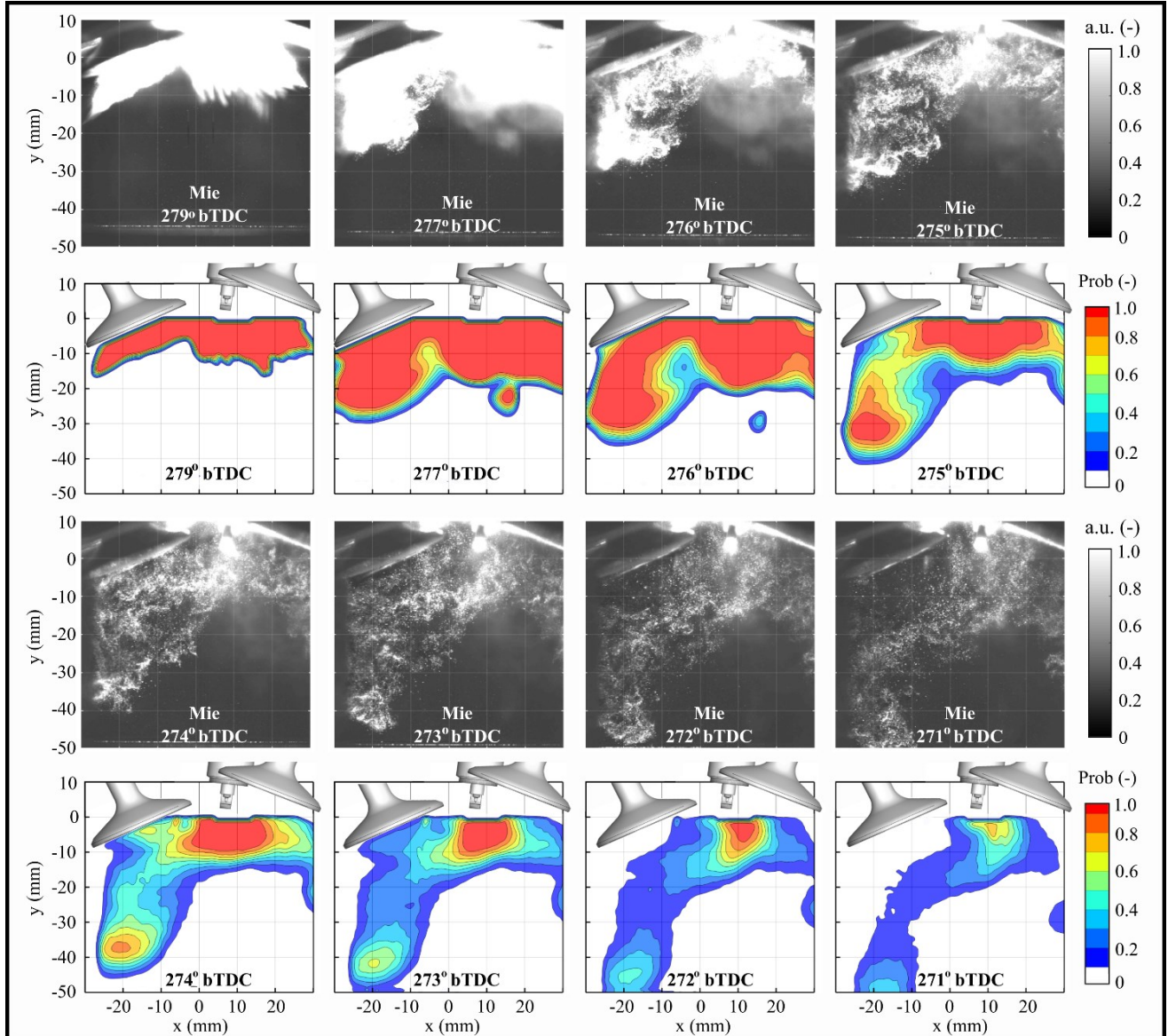

**Figure 2:** Instantaneous Mie scattering images from a single cycle (1$^{st}$ and 3$^{rd}$ rows). Ensemble-cycle 2D fuel droplet PDFs (2$^{nd}$ and 4$^{th}$ rows). Ensemble probabilities based on 288 cycles (fuel droplets characterised by a threshold scattering intensity).

Immediately following start of injection (SOI, 280° bTDC), Mie images from 279°-277° bTDC reveal the influx of liquid fuel during injection. Scattered light off liquid fuel during injection saturates the camera and the spray appears as high-intensity regions in the images. During injection, the hollow-cone spray geometry is quickly distorted as the fuel impacts the dual intake valves located directly beneath the fuel injector. At 277° bTDC liquid fuel penetrates through the narrow channel between the dual intake valves on the left-side of the image, while liquid fuel impacts the spark plug and scatters fuel on the right-side of the image. Signal intensities from multiple-scattering saturate the 12-bit HS-PIV camera (4096 counts) in the fuel spray during injection. Signal intensities are below the saturation limit after 275° bTDC.

After injection, liquid fuel regions break up rather quickly into regions of dispersed fuel droplets. The fuel that passed through the narrow channel between the intake valves progresses downwards as a cluster of droplets and disperses with a spatial propensity towards the left-side of the FOV. As °CA progresses beyond 275° bTDC regions of high probability decrease in size and exhibit progressively lower probabilities, clearly indicating the dispersion and possible evaporation of the liquid fuel. Conglomerated droplet regions near the spark plug appear to disperse more slowly and do not progress as far downwards into the FOV during the image timing.

TPIV particle reconstruction was possible starting at 274º bTDC when particle distributions became progressively sparse. Figure 3 shows the ensemble-average particle per pixel (ppp) fields for TPIV images at selected ºCAs. Figure 3d shows the ppp PDFs for all TPIV ºCAs imaged. Ppp values range from 0.02-0.087 with spray regions having values above 0.04. Ppp values are highest near the top of the TPIV image where fuel droplets are scattered off the intake valve and spark plug. Ppp values are highest at 274º bTDC, for which the PDF shows a bimodal distribution. Ppp values decrease with ºCA as fuel droplets disperse and evaporate. The ppp distribution in the left regions describes the fuel that passed in between the two intake valves. The droplet distribution is less than the top regions because the intake valves block a significant portion of the fuel spray that would otherwise be present on the left side. In this work, the flow field is primarily studied within the left side region where ppp values are lower than the top portion of the TPIV images.

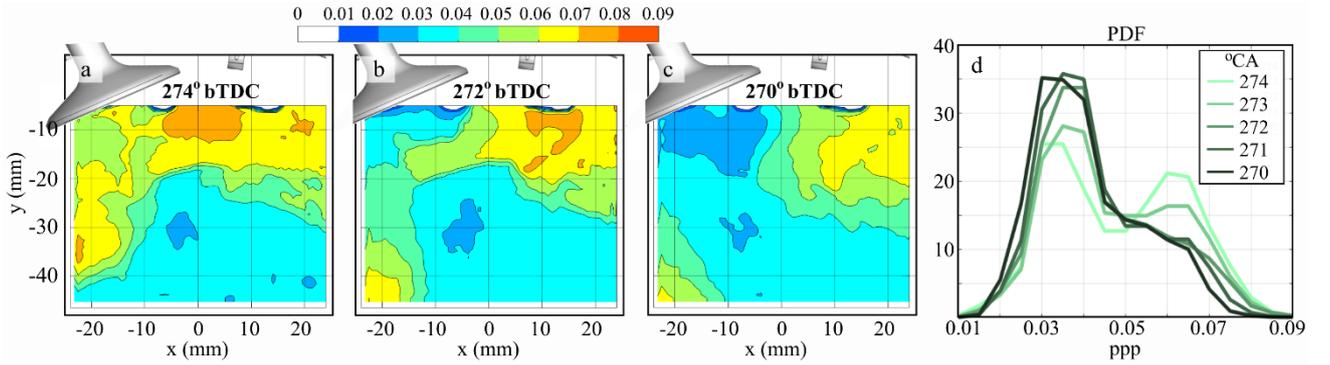

Figure 3: (a-c) Ensemble-average ppp fields for TPIV shown on the z = 0 mm plane for selected ºCAs. (d) PDF of ppp for all TPIV ºCAs. Statistics are based on 100 cycles at each ºCA.

The number of particles (i.e. ppp) and the particle image diameter are important quantities to consider for tomographic reconstruction. For this, the source density, $N_S$, defined in eqn. 1, should be considered (Novara et al. 2010; Scarano 2013):

$$N_S = ppp \cdot \frac{\pi}{4} d_\tau^{*2} \qquad (1)$$

where $d_\tau^{*2}$ is defined as the pixel normalized particle image diameter (Scarano 2013). In this work, average $d_\tau^{*2}$ values were $d_\tau^{*2}$ = 2-3. Accurate tomographic reconstruction is anticipated for $N_S \leq 0.3$ for MART algorithms and $N_S$ should remain below 0.5 in order to avoid loss of optical transmission (Novara et al. 2010; Discetti and Astarita 2012). For $d_\tau^{*2}$ = 2, ppp values as high as 0.085 yield $N_S$ = 0.27, while $d_\tau^{*2}$ = 3 yields $N_S$ = 0.5 when ppp = 0.074. Thus, regions with ppp < 0.074 are anticipated to provide sufficient tomographic reconstruction quality. Areas exceeding ppp > 0.074 are ignored within this work.

## 3.2 Fidelity of PIV particles

Fuel droplets, acting as particle tracers, can influence velocimetry findings. In order to portray gas velocity, the fuel droplets should behave similarly to oil droplets and accurately follow the gas flow. Evaluating the particle response time is required to determine whether a given droplet will accurately follow the gas flow. The droplet response time is calculated as (Tropea et al., 2007):

$$t_0 = \rho_d d_d^2 / 18\mu_g \qquad (2)$$

where $\rho_d$ is the droplet density, $d_d$ is the droplet diameter, and $\mu_g$ is the gas dynamic viscosity. In this work, $\rho_d$ = 690 kg/m³ (isooctane) and $\mu_g$ = 1.85 × 10⁻⁵ Ns/m² (air evaluated at T = 300 K) (Green and Perry 2008). The droplet response time should be lower than the relevant timescales of the engine flow. The turbulent turn-over timescale is considered as a relevant timescale and defined as:

$$\tau_t = L/u' \qquad (3)$$

where $L$ is the length scale of a typical eddy and $u'$ is a representative RMS velocity associated with these eddies. For engine flows, $L$ is often estimated as the 1/6 the height between the piston and cylinder head (approx. 65 mm / 6 = 10.8 mm) (Lumley 1999). This estimation is primarily used to estimate length scales during compression and not during intake. The PIV images in this work suggest typical eddy sizes of 3-5 mm in diameter, which are used to estimate $L$. The $u'$ value is taken as the maximum RMS velocity from Reynolds decomposition, which is determined as $u' = 10$ m/s from TKE images discussed in Sect. 4.1. This yields $\tau_t$ values ranging from 300 – 500 μs.

Determining droplet diameter is best measured using phase Doppler techniques. Unfortunately, such equipment was not available for these measurements. Therefore, in this work we perform an alternative approach using geometric optic assumptions to *approximate* droplet diameters in order to *approximate* $t_0$. This approach, used by Peterson et al., (2017), utilizes a Mie Scattering approach and a particle tracking velocimetry (PTV) algorithm applied to the HS-PIV dataset. The PTV algorithm (LaVision) identified the location of individual particles for which the particle intensity ($I$) was extracted. This analysis was performed for injection and non-injection operation within the entire FOV below y = 0 mm for 288 cycles. Findings from cycles without injection identified oil droplets with average diameter $<d_{oil}> = 0.5$ μm and average intensity $<I_{oil}> = 580$ counts. The individual droplet diameters for injection operation were estimated by:

$$d_{droplet} = <d_{oil}> \cdot (I_{droplet}/<I_{oil}>)^{1/2} \quad (4)$$

The maximum droplet intensity was 3861 counts, which is less than the camera saturation level of 4096 counts. Figure 4 shows $d_{droplet}$ vs. $t_0$ at 274° and 271° bTDC. Particle response times range from $t_0 = 2.0$-14.0 μs, corresponding to droplet diameters (fuel or oil) of $d_{droplet} = 0.9$-2.6 μm. Droplet diameters cover the same range for both °CAs. Response times remain below 15 μs, yielding Stokes numbers, $St = t_0/\tau_t < 0.1$, which indicate that fuel droplets should follow the gas-flow. This is unsurprising since images occur 0.8-1.4 ms after injection in an expired spray plume far from the injector nozzle.

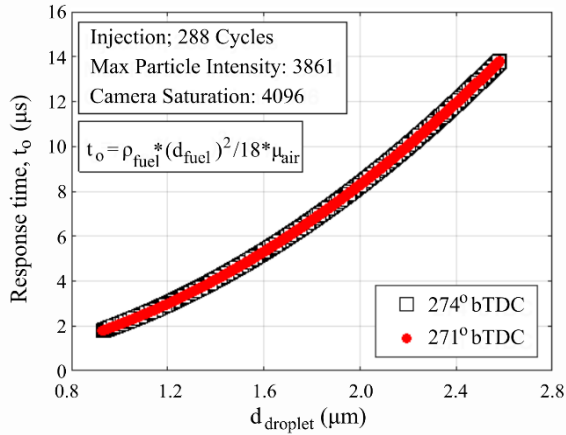

**Figure 4:** Droplet response time vs. droplet diameter for injection operation (288 cycles). Analysis is performed using PTV within the entire FOV below y = 0 mm.

It should be emphasized that this PTV / Mie Scattering method simply provides approximate $d_{droplet}$ and $t_0$ values. Factors such as multi-scattering and particle image overlap can increase the droplet intensity within the Mie scattering images, especially for injection operation. This increased intensity ($I_{droplet}$) would effectively overestimate $d_{droplet}$ and $t_0$. Thus, values reported in Fig. 4 could in fact be conservative. This, however, further strengthens the argument that that fuel droplets are expected to follow the gas-flow. While this method is not intended to measure $d_{droplet}$ with high accuracy, it is suitable for the intended assessment. Furthermore, despite potential inaccuracies in calculating $d_{droplet}$, larger particle sizes up to 9 μm diameter would yield $St = 0.3$. Thus, even if particle diameters are underestimated by a factor of 3-4, the particle response time is still well-below the turbulent timescales.

### 3.3 TPIV assessment from HS-PIV

To further assess the viability of TPIV to resolve the spray-induced flow field, HS-PIV is analyzed alongside TPIV measurements. HS-PIV was utilized to capture the 2D2C velocity field (z = 0 mm) evolution from 285° to 271° bTDC, leading up to a synchronized TPIV image taken at 270° bTDC. This combined set-up, which was employed for injection and non-injection operation, enabled time-correlated measurements between both PIV methods, thus allowing TPIV results (z = 0 mm) to be benchmarked against the well-established HS-PIV technique.

Figure 5 presents the flow evolution from two image sets to describe the instantaneous engine flow without fuel injection (top) and with fuel injection (bottom). HS-PIV images are shown for selected °CAs from 278°-271° bTDC, while the image on the far right of each sequence represents the two-component velocity field captured by TPIV (z = 0 mm) at 270° bTDC. The top row for each image set presents Mie scattering images, while the bottom row shows the corresponding 2D2C velocity field (z = 0 mm) represented by streamlines. The TPIV Mie scattering images are taken from camera 2, which was positioned nearly perpendicularly to the imaging volume. Mie images in Fig. 5 are normalized by the maximum intensity for better visualization since HS-PIV and TPIV images were acquired with different light sources.

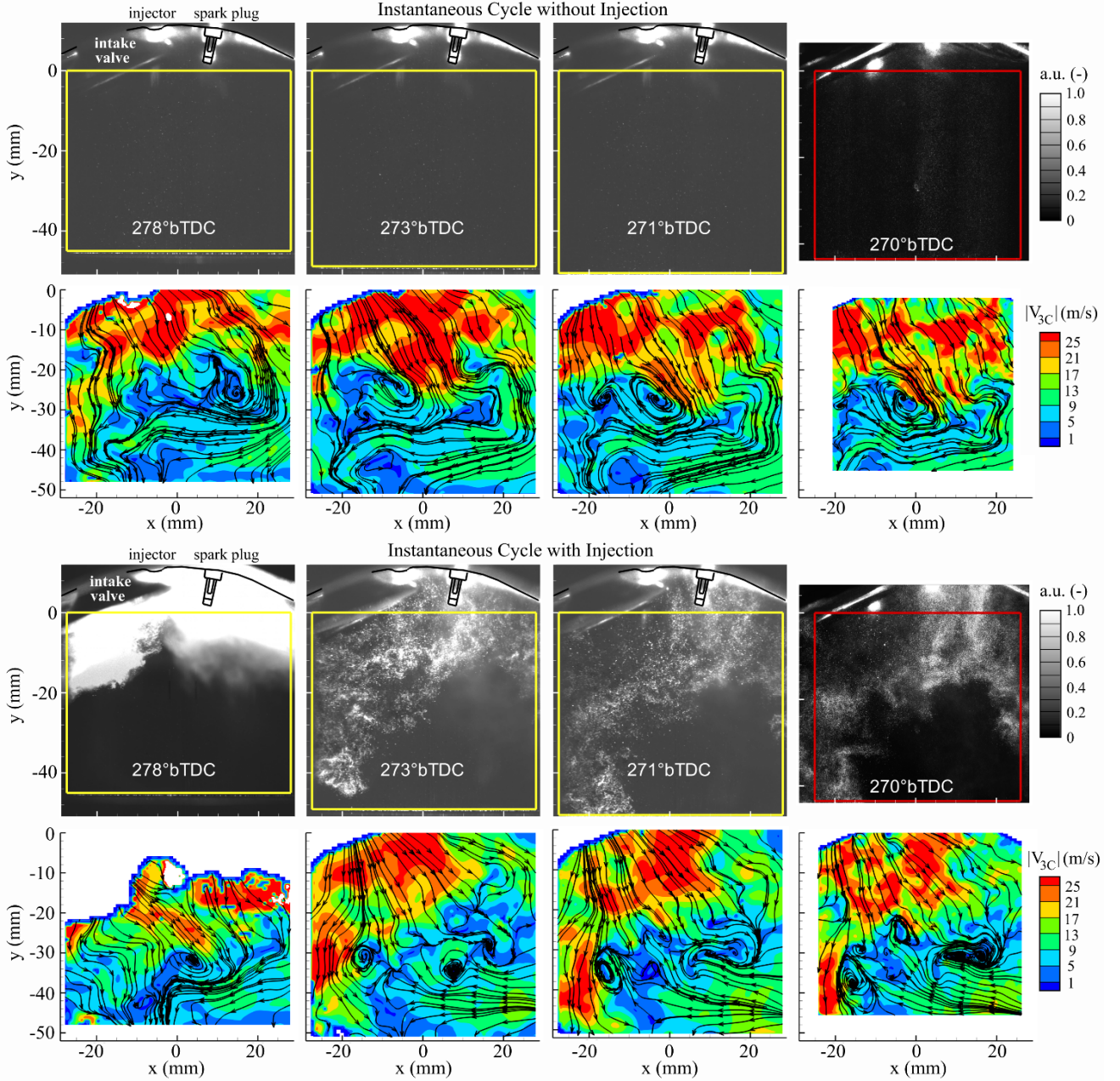

**Figure 5:** Image sequence showing instantaneous Mie and velocimetry images for operation without injection (top) and with injection (bottom). HS-PIV 278-271° bTDC, TPIV (z = 0 mm) 270° bTDC.

Images without injection reveal that velocity distributions are characterized by high velocities entering the cylinder and the downward piston motion. Velocity magnitudes are highest near the intake valves where the annular flow from each intake port impinges on each other, creating a strong jet-like flow into the cylinder. This high-velocity, jet-like flow is referred to as "intake-jet" (Voisine et al. 2011, Freudenhammer et al. 2014). As the flow extends beyond the intake-jet, it is recirculated by the cylinder wall and the piston top, forming a clockwise tumble motion in the current perspective of the symmetry plane. The TPIV image (far right) shows the flow in a smaller FOV and qualitatively shows good agreement with the HS-PIV at the preceding °CA (271° bTDC).

Mie scattering and velocity images with injection show the distribution of fuel droplets and the spray-induced flow field for the same cycle presented in Fig. 2. At 278° bTDC, the obscuring liquid spray regions are masked in the velocity field, but the remaining flow field shows the pre-existing tumble flow formation. Liquid fuel penetrating between the intake valves quickly progresses downwards in the z = 0 mm plane on the left-side of the image with velocities exceeding 25 m/s from 278º-270º bTDC. For simplicity, this high-velocity fuel droplet region that penetrates between the intake valves and progresses through the FOV on the left will be referred to as the "spray-induced jet" (SIJ). From 274º-270º bTDC, the intake-jet is also observed for injection cycles. After injection, fuel droplets are dispersed within the upper-half and left side of the FOV, while air-only flows are located on the lower-right of the FOV. As was the case with the non-injection cycle, the TPIV image with injection remains in good agreement with the preceding HS-PIV image taken at 271º bTDC.

Figure 5 provides a qualitative comparison of the 2D2C flow fields between HS-PIV and TPIV for consecutive ºCAs. However, in order to fully validate the use of TPIV for the injection environment, additional quantitative information in support of the visual observations is required. To achieve this, x- and y-velocity components from HS-PIV at 271º bTDC were extracted at each point in space and subtracted from TPIV velocity components at 270º bTDC. This provides a spatially-distributed velocity difference between HS-PIV and TPIV. For reference, this procedure was also performed for HS-PIV between 272º and 271º bTDC. These operations were performed for 144 cycles (i.e. maximum number of synchronized HS-PIV / TPIV datasets) for operation with and without fuel injection. Differences are evaluated within the z = 0 mm TPIV domain at the HS-PIV spatial resolution.

Figure 6 shows PDFs of the velocity differences between HS-PIV and TPIV. Velocity differences are not expected to always equal zero because data is extracted at different ºCAs and velocity changes with ºCA. All PDFs are centered at zero and show similar distributions. Two interesting findings emerge from Fig. 6. The first is that velocity differences are smaller (i.e. narrower distribution) for cycles with injection than cycles without injection. This implies that TPIV is in better agreement with HS-PIV measurements for injection cycles. The second is that velocity differences between HS-PIV and TPIV from 271º-270º bTDC are smaller than HS-PIV differences from 272º-271º bTDC. This holds true for operation with and without injection. This aspect likely results from decreasing in-cylinder velocity magnitudes due to the gradual deceleration of the piston or intake flow phasing at the particular imaging time-frame. Findings presented in Fig. 6 reveal that TPIV is as reliable as the HS-PIV measurements and validates TPIV for the early-injection engine environment performed within this study.

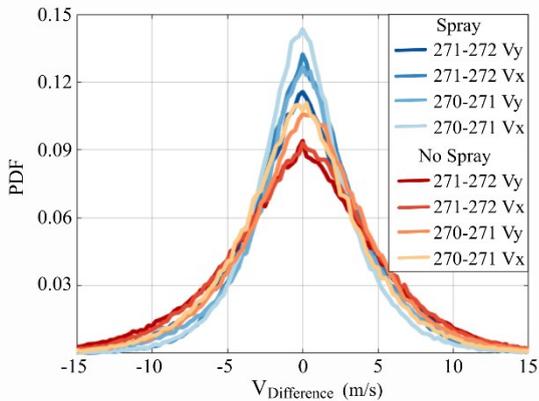

**Figure 6:** PDF of HS-PIV / TPIV velocity differences

## 3.4 Measurement uncertainty

### 3.4.1 Velocity components

Planar PIV and TPIV experimental parameters and processing procedures were optimized to reduce measurement uncertainty. For image optimization, camera and lenses operated with a depth of focus larger than the imaging volume to provide focused particles with high signal intensity. All cameras focused on regions near the cylinder axis to prevent distortion due to the curved glass cylinder (Reuss et al., 2002). TPIV camera angles were chosen to provide the maximum range of camera angles suitable for the FOV. The fuel spray is dilute (3.6 mg/cycle), in order to keep particle distributions at a manageable level. For TPIV, sufficient tomographic reconstruction is anticipated for regions with ppp ≤ 0.074 and regions with ppp > 0.074 were excluded from analyses. The reconstructed signal-to-noise (SNR) level for TPIV, defined as the ratio of reconstructed intensity of the particle region to that of the ghost level intensity (Scarano 2013), remained above 2.3. The volume self-calibration method provided remaining pixel disparities less than 0.2 pixels. Provided the aforementioned particle image qualities, the expected uncertainty in the local average pixel displacement is 2-3%.

For data processing, the multi-pass decreasing window size cross-correlation algorithm used in this work provides a window shift that adaptively improves vector computation. The peak ratio factor was set to 1.5 and a local median filter was used to remove spurious vectors (Westerweel 1994). Average correlation coefficients associated with cross-correlation algorithms ranged from 0.55-0.75 (Westerwheel 1997; Elsinga et al., 2006). The uncertainty calculations in DaVis for 2D PIV yielded velocity component uncertainties within $U_u$ = 4% and $U_v$ = 4%.

Uncertainty calculations are not currently available in DaVis for TPIV. As shown in Sect. 3.3, x- and y-velocity components from TPIV are in good agreement with PIV. Thus, it is anticipated that $U_u$ and $U_v$ values associated with TPIV will be similar to PIV. Moreover, TPIV SNR levels and source density values are considered sufficient for particle reconstruction (Novara et al. 2010, Michaelis et al. 2010). Thus, it is not expected to have significantly larger uncertainties associated with TPIV. The *estimated* uncertainty associated with TPIV is reported to be: $U_u$ = 5%, $U_v$ = 5% and $U_w$ = 8%. The slightly larger $U_w$ values are anticipated due to the limited number of camera views for particle reconstruction in the z-direction. Justification of these uncertainty values is discussed in Sect. 3.4.2.

3.4.2 Velocity divergence (TPIV)

The propagation of uncertainty principle (Coleman and Steele 2009; Sciacchitano and Wieneke 2016) is used to calculate the uncertainty associated with quantities derived from velocity components. In this section, this principle is applied to the velocity divergence term, where the calculated uncertainties can be compared to the deviation from the conservation of mass. This analysis is performed for non-injection operation, when density is considered uniform. The propagation of uncertainty analysis is further applied to individual vorticity and strain rate components for injection operation in Sect. 3.4.3.

The propagation of uncertainty principle considers a derived quantity of interest $q$ that is a function $F$ of $N$ measured variables $\lambda_i$ with $i$ = 1,2, ..., $N$.

$$q = F(\lambda_1, \lambda_2, ..., N) \qquad (5)$$

When quantity $q$ is derived from a single measurement, much like vorticity or divergence, Coleman and Steele (2009) demonstrate that the uncertainty $U_q$ of $q$ can be represented by the propagation of uncertainty principle:

$$U_q^2 = \sum_{i=1}^{N}\left(\frac{\partial F}{\partial \lambda_i}\right)^2 U_{\lambda_i}^2 + 2\sum_{i=1}^{N-1}\sum_{j=i+1}^{N}\frac{\partial F}{\partial \lambda_i}\frac{\partial F}{\partial \lambda_j}\rho\left(\delta_{\lambda_i},\delta_{\lambda_j}\right)U_{\lambda_i}U_{\lambda_j} \qquad (6)$$

Where $\rho\left(\delta_{\lambda_i},\delta_{\lambda_j}\right)$ is the cross-correlation coefficient between the uncertainty of $\lambda_i$ and $\lambda_j$, which are indicated by $\delta_{\lambda_i}$ and $\delta_{\lambda_j}$, respectively. For PIV, the $\rho\left(\delta_{\lambda_i},\delta_{\lambda_j}\right)$ term is a function of the percentage overlap used during PIV processing (Sciacchitano and Wieneke 2016).

In this work, spatial velocity gradients are derived by the first-order central difference scheme. Each velocity component is considered as discrete functions, defined by grid points with uniform grid spacing, $d$ = 0.375mm, in the $x$, $y$, and $z$ direction. Thus, the velocity divergence term is calculated as:

$$DIV(x,y,z) = \frac{1}{2d}[(u(x+d,y,z) - u(x-d,y,z)) + (v(x,y+d,z) - v(x,y-d,z)) \\ + (w(x,y,z+d) - w(x,y,z-d))] \qquad (7)$$

Using the propagation of uncertainty principle (eqn. 6), Sciacchitano and Wieneke (2016) demonstrate that the local uncertainty $U$ associated with the velocity divergence term is:

$$U_{DIV}(x,y,z) = \frac{1}{d\sqrt{2}}\sqrt{(1-2\rho d)(U_u(x,y,z)^2 + U_v(x,y,z)^2 + U_w(x,y,z)^2)} \qquad (8)$$

where $2\rho d$ has a value of 0.45, which is associated with 75% overlap in TPIV processing (Sciacchitano and Wieneke 2016). The local uncertainty of individual velocity components is determined by multiplying the velocity component value by its uncertainty, i.e.:

$$U_{u_i}(x,y,z) = u_i(x,y,z) * U_{u_i} \qquad (9)$$

$U_{DIV}$ is calculated for each $x,y,z$ location within the imaging volume using eqn. 8. $U_{DIV}$ is then normalized by the local, absolute value of the velocity divergence to provide a relative uncertainty, $U_{DIV}/|DIV|$.

The velocity divergence can be evaluated directly from TPIV data to assess the deviation from conservation of mass:

$$\rho^{-1}(\partial\rho/\partial t) + \partial u_i/\partial x_i = 0 \tag{10}$$

During intake, without injection, the first term in eqn. 10 can be neglected, yielding:

$$DIV = \partial u_i/\partial x_i = 0 \tag{11}$$

Within the literature, a relative deviation from zero divergence is often used to assess TPIV uncertainty (e.g. Baum et al. 2013; Cortion et al. 2014; van Overbrueggen et al. 2015). In this work, we do not report the deviation from zero divergence as the final TPIV uncertainty. Instead, we assess this relative deviation to better understand the relative uncertainty values, $U_{DIV}/|DIV|$.

The deviation from zero divergence is assessed similar to Baum et al. (2013). The velocity difference ($\Delta U = \Delta u + \Delta v + \Delta w$) is calculated for cubic control volumes (CV) of equidistant grid-spacing (0.375 mm) throughout the entire measurement volume. To quantify the relative deviation from zero divergence, $\Delta U$ is normalized by the averaged velocity ($|V|_{3D,CV}$) that enters each CV. Figure 7 shows the PDF of $\Delta U / |V|_{3D,CV}$. For brevity, this is only shown for 270° bTDC, but other °CAs are similar for non-injection operation. This PDF represents a normal distribution symmetric around zero. Twice the standard deviation, $2\sigma = 0.24$ (i.e. 24%), reports the relative deviation from zero.

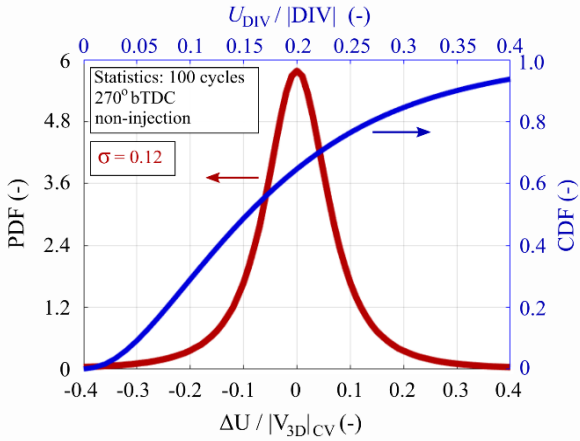

**Figure 7:** PDF of $\Delta U / |V|_{3D,CV}$ and CDF of $U_{DIV}/|DIV|$. Statistics are based on 100 cycles for non-injection operation. Data is shown for 270° bTDC.

The PDF of $U_{DIV}/|DIV|$ (not shown) resembles a chi-distribution. For this distribution, the cumulative distribution function (CDF) is used to evaluate the probability that the uncertainty will be less than or equal to a given value. Figure 7 shows the CDF of $U_{DIV}/|DIV|$ for 270° bTDC. Using $2\sigma = 0.24$ from the PDF of $\Delta U / |V|_{3D,CV}$ as guidance, it is shown that $U_{DIV}/|DIV| = 0.24$ at CDF = 0.75.

The PDF of $\Delta U / |V|_{3D,CV}$ shows that 95% of the data (i.e. $2\sigma$) is within 24% of an accepted value, while the CDF of $U_{DIV}/|DIV|$ shows that only 75% of the data contains uncertainty values lower than 24%. A couple of points should be made about this observation. (1) The PDF and CDF in Fig. 7 are different distributions of two different variables; the probability of data within a given range is not expected to be exactly the same. (2) $U_{DIV}$ is calculated using $U_u = 5\%$, $U_v = 5\%$ and $U_w = 8\%$. It is expected that $U_{u_i}$ uncertainties are *within* these limits, while $U_{DIV}$ in eqn. 8 is calculated as if $U_{u_i}$ are *exactly* these values. Thus, $U_{DIV}$ may be slightly overestimated in some cases, such that a lower percentage of data falls within a given uncertainty value. (3) The minor discrepancies between PDF and CDF findings suggest the reported $U_{u_i}$ values are reasonable estimates. This is important since these uncertainties were not directly measured and are used within Sect. 3.4.3.

Acknowledging these findings, the CDF curves are used to report reasonable uncertainty values for quantities derived from velocity components. In this manuscript, uncertainty is reported where CDF = 0.8. While this value is arbitrary, it provides a conservative, yet reasonable uncertainty estimation. This criterion is used to evaluate vorticity and strain rate uncertainty for injection operation in Sect. 3.4.3.

3.4.3 Vorticity and strain rate uncertainty (TPIV)

The propagation of uncertainty (eqn. 6) is further used to calculate uncertainty associated with vorticity and strain rate components. Unlike the divergence analysis, this uncertainty analysis is performed for cycles with fuel injection. Using the propagation of uncertainty, Sciacchitano and Wieneke (2016) demonstrate that the local uncertainty associated with individual vorticity and strain rate components is:

$$U_{\Omega_k, S_{ij}}(x,y,z) = \frac{1}{d\sqrt{2}}\sqrt{(1-2\rho d)\left(U_{u_i}(x,y,z)^2 + U_{u_j}(x,y,z)^2\right)} \qquad (12)$$

Since this work is concerned with evaluating high turbulence levels associated with the spray, uncertainties are calculated for locations where the absolute value of $\Omega_k$ or $S_{ij}$ exceed 6,000 s$^{-1}$. Section 4.4 shows that local $\Omega_k$ and $S_{ij}$ magnitudes associated with the spray-induced turbulence are in the range of 6,000 to 15,000 s$^{-1}$. Similar to the divergence analysis, $U_{\Omega_k, S_{ij}}(x,y,z)$ is normalized by the absolute value of $\Omega_k, S_{ij}$ to provide a relative uncertainty.

The CDFs of the relative uncertainties are shown in Fig. 8a,b. These distributions are based on 100 cycles at 274° bTDC. Table 3 reports these uncertainties at CDF = 0.8 for each quantity. Uncertainties remain below 21%, while $\Omega_x$ and $S_{yz}$ exhibit the largest uncertainties. This is due to higher uncertainties associated with $U_w$ and the fact that y-velocities (i.e. $v$) are typically the largest within the imaging volume, which increases the $U_{u_j}(x,y,z)$ term in eqn. 12.

Figure 8c,d shows that the uncertainty decreases with °CA. For brevity, this is only shown for $\Omega_x$ and $S_{yz}$ (i.e. quantities with the largest uncertainty), but trends are consistent for all quantities. The uncertainty decreases due to the reduction of velocity magnitude after injection. This velocity reduction represents the decay of spray-induced turbulence, which is discussed further in Sect. 4. Table 3 reports quantity uncertainties at 270° bTDC to quantify the decrease in uncertainties at the latest °CA imaged.

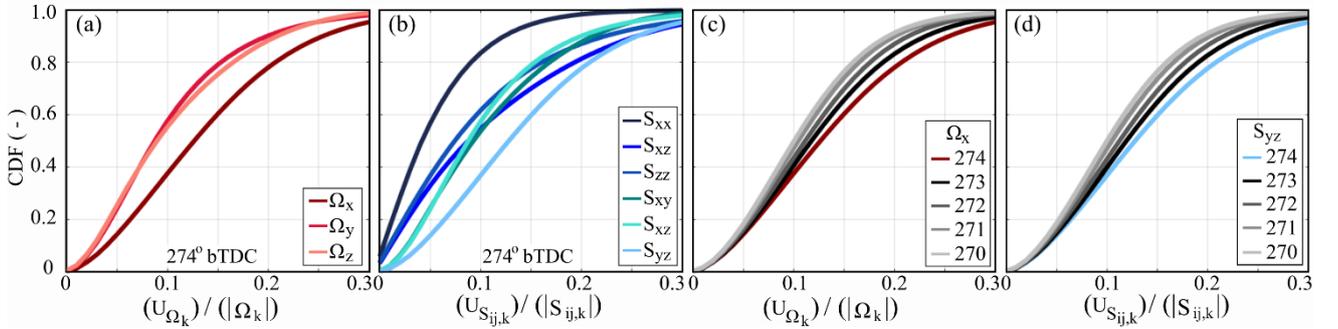

**Figure 8:** CDF of normalized uncertainty associated with individual vorticity and strain rate values. (a,b) CDF values at 274° bTDC. (c,d) CDFs for selected components to show the decrease of uncertainty with °CA.

*Table 3: Uncertainty Analysis*

| Variables in eqn. 12 | | |
|---|---|---|
| $U_u$ = 5%; 0.05 | $U_v$ = 5%; 0.05 | $U_w$ = 8%; 0.08 |
| $2\rho d$ = 0.45 | $d$ = 3.75x10$^{-4}$ m | |
| Uncertainty (CDF = 0.8), 274° bTDC | | |
| $U_{\Omega_x}$ = 20.7% | $U_{\Omega_y}$ = 15.3% | $U_{\Omega_z}$ = 16.5% |
| $U_{S_{xx}}$ = 9.0% | $U_{S_{yy}}$ = 19.3% | $U_{S_{zz}}$ = 16.0% |
| $U_{S_{xy}}$ = 16.8% | $U_{S_{xz}}$ = 15.8% | $U_{S_{yz}}$ = 20.7% |
| Uncertainty (CDF = 0.8), 270° bTDC | | |
| $U_{\Omega_x}$ = 15.8% | $U_{\Omega_y}$ = 12.8% | $U_{\Omega_z}$ = 13.4% |
| $U_{S_{xx}}$ = 8.6% | $U_{S_{yy}}$ = 14.4% | $U_{S_{zz}}$ = 13.7% |
| $U_{S_{xy}}$ = 12.8% | $U_{S_{xz}}$ = 13.6% | $U_{S_{yz}}$ = 15.8% |

Figure 9 shows joint PDFs to describe the uncertainty distribution in more detail. For brevity, this is performed for the quantity $\Omega_x$ because it exhibits the largest uncertainty, but all other quantities show similar findings. Figure 9a shows the joint PDF of $\Omega_x$ and normalized uncertainty, while Fig. 9b shows the joint PDF of y-velocity ($v$) and normalized uncertainty. Figure 9a shows that higher uncertainties are associated with lower $\Omega_x$ magnitudes and tend to decrease within increasing $|\Omega_x|$. In Sect. 4, it is shown that higher turbulence levels associated with the spray exhibit $|\Omega_k|$ in excess of 12,000 s$^{-1}$. Thus, uncertainties reported in Table 3 may be considered conservative estimates for spray-induced turbulence regions. Figure 9b shows that uncertainty values tend to increase with increasing $v$ magnitudes. This can intuitively be seen from eqn. 12 and is only true for quantities containing derivatives of *v*. This also explains why quantities involving *v* typically have higher uncertainties.

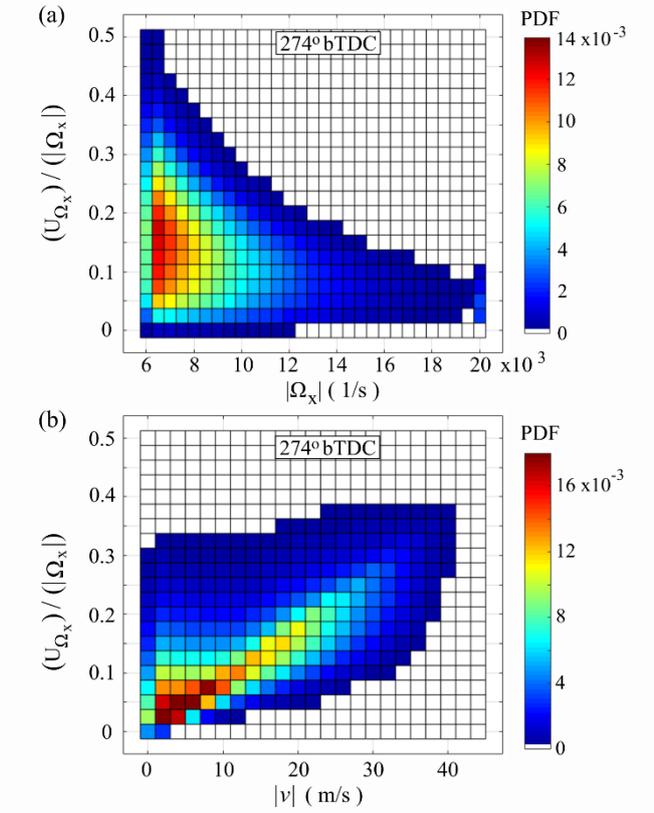

**Figure 9:** Joint PDF of (a) $|\Omega_x|$ and normalized uncertainty and (b) y-velocity (*v*) and normalized uncertainty. Uncertainty decreases with increasing $|\Omega_x|$ and increases with increasing *v*.

## 4. Results

### 4.1 3D ensemble-average and turbulent kinetic energy distributions

The 3D ensemble-average velocity and turbulent kinetic energy (TKE) are evaluated within the volumetric domain to describe the spray-induced flow field behavior and identify locations of intense turbulence in relation to spray-induced flow features. Figure 10 shows 3D isosurfaces of ensemble-average velocity magnitude and TKE at selected °CAs after injection. Streamlines representing the ensemble-average flow-field are shown within the z = 0 mm plane. For comparison, the ensemble-average velocity and TKE distributions for non-injection operation is also shown in Fig. 10. For brevity, the non-injection distributions are only shown at 270° bTDC since the spatial distributions are similar from 274°-270° bTDC. All distributions are based on phased-locked TPIV images for 100 engine cycles. 3D TKE is calculated as:

$$TKE = (1/2)\overline{u'_i u'_i} \qquad (13)$$

where $u'_i$ is the fluctuating velocity component in the $i^{th}$ direction. TKE is calculated by Reynolds decomposition.

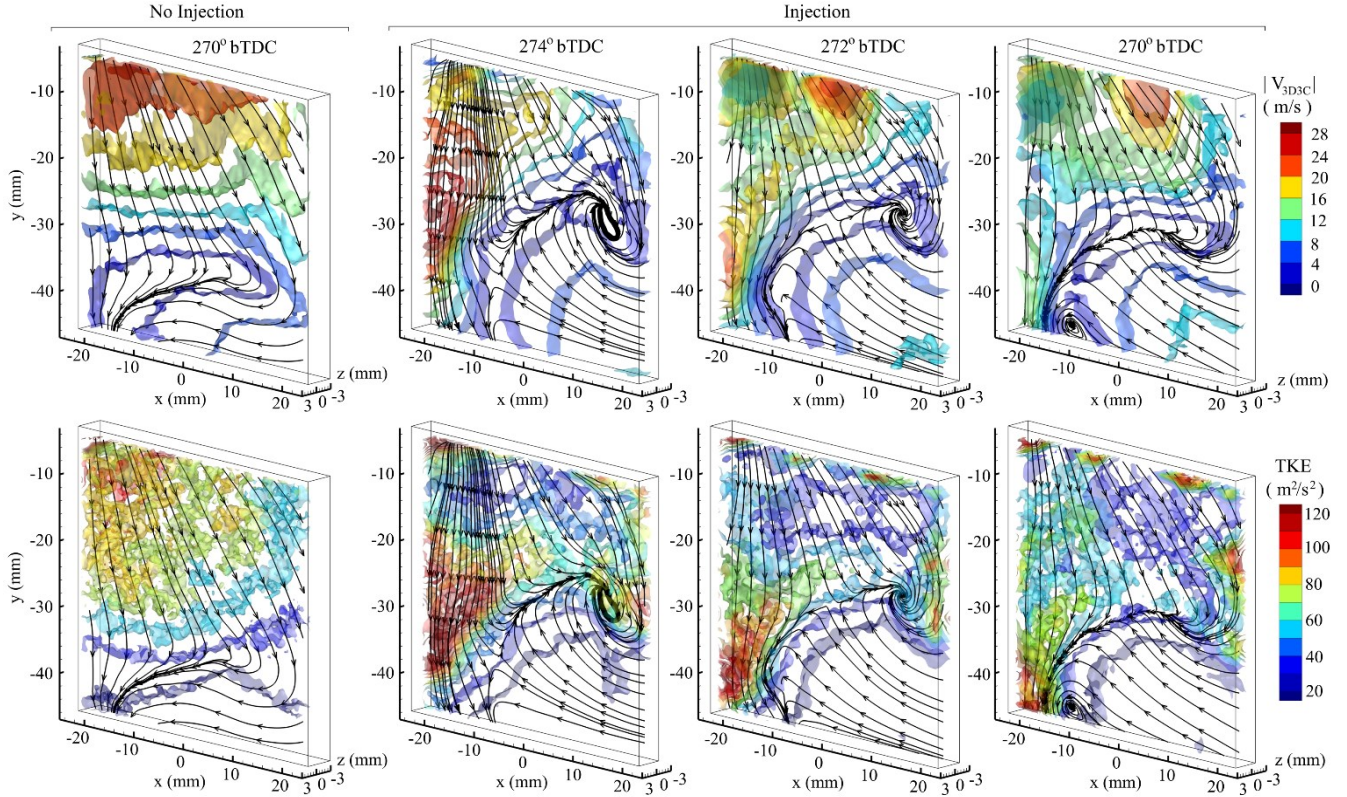

**Figure 10:** 3D iso-surfaces of ensemble-average velocity magnitude (top) and turbulent kinetic energy (bottom). Images shown for selected °CA with and without injection. Statistics are based on 100 images at each °CA. Ensemble-average flow is represented by streamlines shown in the z = 0 mm plane.

For non-injection operation, intake air velocities are largest within the intake-jet region (upper left) and decrease monotonically along the −y direction. TKE values are also largest within the intake-jet region, spanning a range from 70-100 $m^2/s^2$. Lowest velocity and TKE values exist near the tumble center. The turbulent flow characteristics of the intake-jet and tumble vortex without injection have been analyzed in detail by Zentgraf et al. (2016).

Velocity and TKE distributions are quite different for injection operation. After injection, the flow field shows the high-velocity SIJ region penetrating downwards on the left side, while a pronounced clockwise vortex exits on the right side of the imaging volume. This vortex is a combination of the generating tumble flow and the spray-induced toroidal flow typically formed at the edge of the hollow-cone spray. As °CA progresses towards 270° bTDC the SIJ progresses through the imaging volume and a counter-clockwise rotating vortex is formed along the periphery of the SIJ near the bottom of the imaging volume. TKE distributions show that turbulence is greatest within the vicinity of the SIJ and is much larger than TKE values associated with any other flow feature including the intake-jet for non-injection cycles. The SIJ is unique to the injection operation and represents a turbulent-infused fuel-air jet that imposes turbulence onto the nearby air flow. The SIJ therefore presents itself as an intriguing flow feature to study the spray-induced flow physics associated with direct-injection, and as such, forms the central focus of the work presented hereafter.

For completion, a comment should be made about the intake-jet region for injection operation. While the general flow direction of the intake-jet is similar between injection and non-injection operation, velocity magnitudes and TKE values are lower for injection operation. The intake-jet without injection exhibits high TKE originating from unsteady turbulent flow behavior (e.g. flow separation, vortex shedding) as well as cyclic variances of the location and direction of mean flow features (Zentgraf et al. 2016). It appears that fuel injection modifies the turbulent behavior observed in the intake-jet region for the °CAs shown. However, detailed analysis of the intake-jet and spray is not the focus of this study.

### 4.2 3D3C velocity distributions and SIJ characterisation

The SIJ is defined as the high-velocity jet progressing downwards on the left-side of the FOV. In this work, the SIJ is characterized by velocity magnitude. A rectangular study volume is used to sample the 3C velocity magnitudes ($|V|_{3C}$) to obtain a consistent definition of the SIJ amongst all cycles and °CAs. Figure 11 illustrates the spatial domain of the study volume. The location and dimensions of the study volume were chosen to isolate the spatial region where the SIJ was observed. The study volume is defined with a height of 20 mm, extending from the bottom edge of the FOV (y = -45 mm) up to y = -25 mm. The height was not extended further into the y-direction because this would increase the likelihood of sampling velocities associated with the intake-jet instead of isolating velocities associated with the SIJ. The study volume is bounded along the x-direction from -25 to -10 mm and z-direction from -2.0 to 2.0 mm.

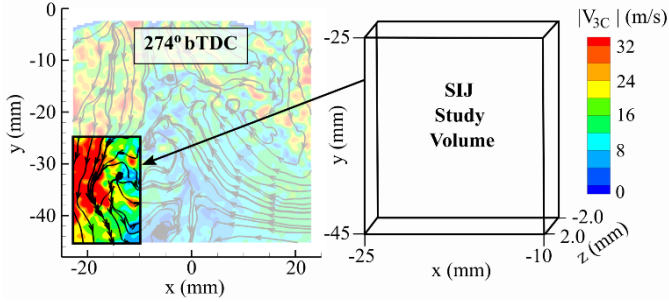

**Figure 11:** Example velocity field (z = 0 mm) highlighting the spatial location of the SIJ study volume where the spray-induced turbulence is further analyzed.

Figure 12 shows $|V|_{3C}$ PDFs for operation with and without injection from 274°-270° bTDC. PDFs include the entire study volume for 100 cycles at each °CA. The PDFs from non-injection operation exhibit a similar unimodal distribution with a mode between 5-7 m/s and a long tail towards velocities up to 30 m/s. This tail is associated with peripheral regions of the intake-jet protruding into the study volume. For injection operation, the PDFs exhibit a broadened and bimodal distribution where the mode associated with higher velocities is clearly affiliated with the SIJ. Velocity magnitudes are largest at earlier °CAs closer to EOI (i.e. 274° and 273° bTDC). As °CA progresses (i.e. 272°-270° bTDC), velocity magnitudes decrease as the momentum of the spray dissipates as time progresses after EOI.

The velocity distributions presented in Fig. 12 are used to select a velocity threshold to define the SIJ domain at each °CA. At 274° bTDC, velocity magnitudes are the largest; 21.5% of the velocity distribution for injection operation exceed 30 m/s, while this threshold is almost never exceeded for non-injection operation (0.17% probability). These velocities are clearly associated with the SIJ and the velocity threshold ($|V|_{SIJ,274}$) of 30 m/s is used to define the SIJ at 274° bTDC. As °CA progresses onwards from 274° bTDC, a new velocity threshold must be chosen for a consistent definition of the SIJ as spray-induced velocities decrease. The SIJ threshold velocity for subsequent °CA is reduced proportionally to the decrease in ensemble-average velocity magnitude within the 3D study volume. This simple relationship is expressed as:

$$|V|_{SIJ,CA-1} = |V|_{SIJ,CA} \cdot \frac{Mean\ Vel_{CA-1}}{Mean\ Vel_{CA}} \tag{14}$$

The resulting SIJ threshold velocities at each °CA are shown in Fig. 12. This methodology is a simplistic approach that identifies the highest velocity regions induced from injection and are used to define the SIJ. As shown in Fig. 12, the reduction in velocity threshold increases the overlap between velocities associated with injection and those observed without injection. The largest overlap occurs at 270° bTDC, where the SIJ velocity threshold is $|V|_{SIJ,270}$ = 21.2 m/s. For operation without injection only 4.2% of the velocity distribution exceeds the 21.2 m/s threshold. Although this overlap is unavoidable as spray-induced velocities decrease, the overlap at 270° bTDC is small; statistical analysis suggests that ~95% of the velocities exceeding this threshold are attributed to injection. Hence, it is argued that this simplistic approach is suitable to identify the high velocities associated with injection, which define the spatial realm of the SIJ.

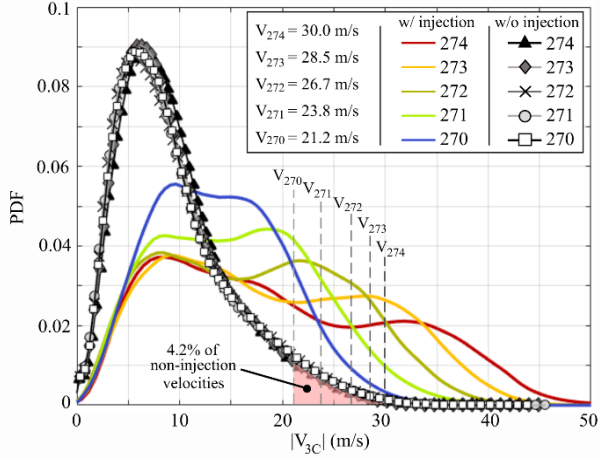

**Figure 12:** PDFs of 3C velocity magnitude ($|V_{3C}|$) within the SIJ study volume for operation with and without injection. Statistics are based on 100 cycles at each °CA.

### 4.3 Spray-induced Strain Rate (S) and Vorticity (Ω) Distributions

#### 4.3.1 Instantaneous velocity, $||S||$ and $||\Omega||$ fields

Instantaneous phase-locked TPIV measurements are utilized to spatially resolve all components of the strain rate (S) and vorticity (Ω) tensors. The complete S and Ω tensors enable quantitative measurements of spatially coherent 3D vortical structures produced from 3D spray-induced shear layers. In this work, distinct regions of high S and Ω are interpreted as regions of high turbulence. Therefore, analysis of the S and Ω distributions is performed to study the spray-induced turbulence associated with the penetrating SIJ flow. Strain rate and vorticity magnitudes are calculated by the Frobenius norm represented by $||\dots||$ and calculated as:

$$||S|| = \sqrt{\sum_{i=1}^{k}\sum_{j=1}^{k} S_{ij}} \qquad (15)$$

$$||\Omega|| = \sqrt{\sum_{i=1}^{k}\sum_{j=1}^{k} \Omega_k} \qquad (16)$$

where

$$S_{ij} = 1/2\left(\partial u_i/\partial x_j + \partial u_j/\partial x_i\right) \qquad (17)$$

and

$$\Omega_k = 1/2\left(\partial u_j/\partial x_i - \partial u_i/\partial x_j\right) \qquad (18)$$

Figure 13 presents instantaneous images of velocity, S, and Ω for a single-cycle at 274° bTDC. The images are chosen to describe the prominent turbulent features associated with the spray-induced jet. For clarity, distributions are visualized on the z = 0 mm plane (2D), but are calculated from the 3D3C velocity field. The top row of images presents quantities within the entire 2D FOV, while the middle row presents zoomed images which highlight the SIJ study volume only. The bottom row shows profiles of each quantity extracted along the x-direction at a given y-location. The velocity threshold criterion, discussed in Sect. 4.2, is used to identify the spatial location of the SIJ. The SIJ boundary is shown within the zoomed images (black-lines). For appropriate evaluation of the flow images, the droplet distribution calculated from the 3D particle reconstruction (FastMART) is shown in Figs. 13 d,h. The FastMART image represents the entire 3D particle field within the illuminated volume (~ 4 mm thickness). The particle distribution is further shown in Fig. 13i, which shows the ppp distribution within the SIJ study window for the z = 0 mm plane.

The spray-induced jet, where $|V_{3C}| \geq 30$ m/s, is clearly identified in Fig. 13 and exists as several isolated pockets upstream a larger "U-shaped" region. In this analysis, we focus on this larger U-shaped region of the SIJ. The SIJ exhibits high velocities up to 50 m/s, while the in-cylinder velocities to the right of the SIJ are significantly lower (10-15 m/s). As a result, steep velocity gradients exist along the jet boarder to form what is commonly referred to as "shear layers" in jet-like flows (Bellan 2000). These shear layers are visualized in Fig. 13f as elongated regions with $\|S\| > 3\times10^4$ s$^{-1}$ along the right boundary of the jet and are comprised of the highest $\|S\|$ values within the entire FOV.

Vorticity values are also large along the SIJ periphery. Unlike strain rate, however, regions of largest vorticity values do not crowd the boarder of the SIJ. Instead regions of high vorticity appear as isolated pockets with $\|\Omega\| > 2.5 \times 10^4$ s$^{-1}$ located near the SIJ boarder. The occurrence of these vortical flow regions, and their relation to the shear flow, is discussed further in Sect. 4.4.

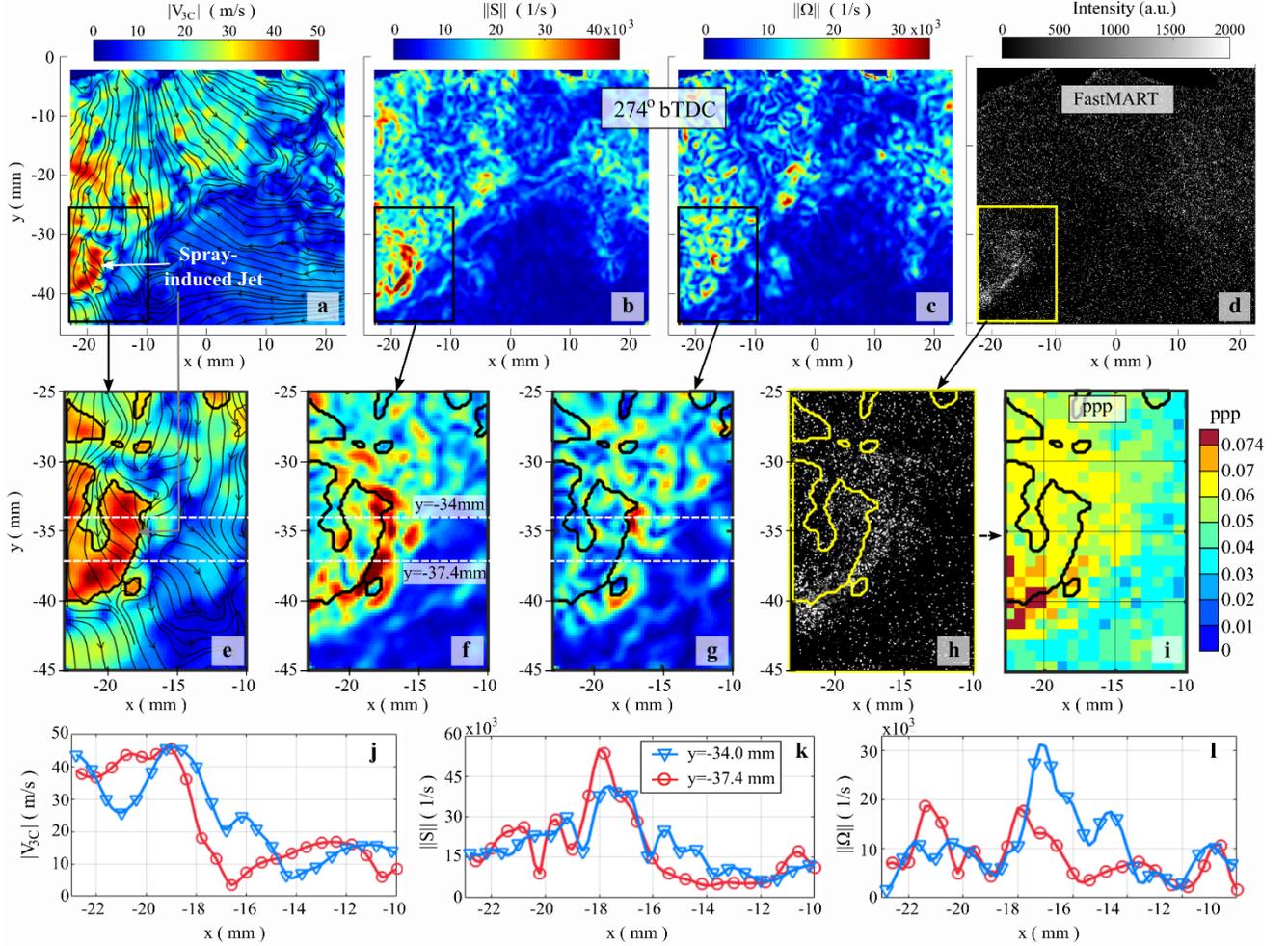

**Figure 13:** Velocity, strain rate, vorticity and FastMART images for an individual cycle at 274º bTDC. FastMART images represent the particle distribution within the entire imaging volume thickness, while all other images show quantity fields on the z = 0 mm plane. The ppp distribution (z = 0 mm) is shown for the FastMART in the SIJ window. Sub-plots j-l show flow quantities extracted along horizontal lines to elucidate the high velocities and sharp velocity gradients associated with the SIJ.

The ppp distribution in Fig. 13i identifies regions of high particle density. Regions with ppp > 0.074 are identified and correspond to discrete areas of concentrated particle distributions in the FastMART images. These regions are located near the bottom of the SIJ and are not associated with regions of high $\|S\|$ and $\|\Omega\|$ along the right side of the SIJ. Regions with ppp > 0.074 are removed from any further analysis presented in this manuscript. The remaining ppp distributions primarily remain between 0.03-0.07 and are considered suitable for tomographic reconstruction. Particle density is greatest at 274º bTDC because it is the closest TPIV timing after injection. Thus, the images shown in Fig. 13 are representative of some of the densest particle distributions evaluated using TPIV.

Figure 14 presents instantaneous images of velocity, S, and $\Omega$ for another single-cycle, but this time at 272º bTDC to demonstrate velocity findings at a later ºCA when local particle distributions are less dense. The layout of images in Fig. 14 is similar to that presented in Fig. 13, but quantity profiles are not shown for brevity. The ppp distribution in Fig. 14i shows a FastMART particle distribution with much lower particle concentrations compared to 274º bTDC. All ppp values are below 0.07 and the majority of the SIJ periphery exhibits ppp values between 0.04-0.06. Although the particle distribution is significantly lower at 272º bTDC, turbulence characteristics surrounding the spray-induced jet are very similar to those shown at 274° bTDC. Namely, high ||S|| and ||Ω|| exist along the periphery of the SIJ. The ||S|| and ||Ω|| surrounding the SIJ remain to be the highest observed in the entire FOV. The qualitative similarities between Figs. 13 and 14 suggests that high ||S|| and ||Ω|| values are not considered to be an artificial result from erroneous tomographic reconstruction due to dense particle distributions. Instead, it is suggested that these regions are associated with strong shear layers and coherent vortical structures formed as the SIJ penetrates through lower velocity regions.

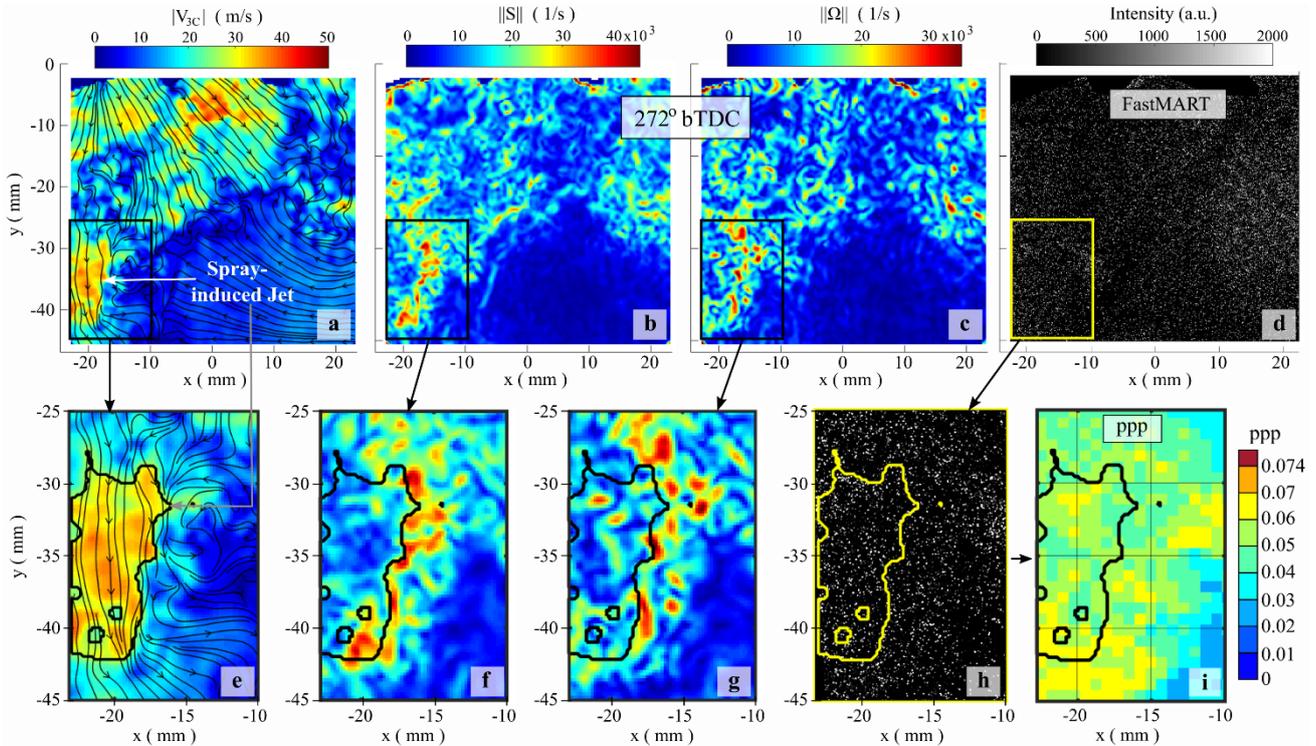

**Figure 14:** Velocity, strain rate, vorticity and FastMART images for an individual cycle at 272° bTDC. FastMART images represent the particle distribution within the entire imaging volume thickness, while all other images show quantity fields on the z = 0 mm plane.

### 4.3.2 Statistical analysis of spray-induced ||S|| and ||Ω||

A statistical analysis is presented to quantify the spray-induced turbulence trends for all cycles. This analysis is intended to elucidate the qualitative trends shown in Sect. 4.3.1 and to further understand the spatial extent of the turbulence into the surrounding in-cylinder flow field. For this analysis, ||S|| and ||Ω|| are extracted from selected zones (i.e. spatial domains) relative to the SIJ volume. These zones exist as layered regions extending perpendicularly in 1 mm increments from the SIJ boundary. An example of these spatial domains is shown in Fig. 15 for an individual cycle at 274° bTDC. This analysis is performed for each z-plane to construct a 3D layer within the imaging volume. For comparison, ||S|| and ||Ω|| are also extracted within the SIJ spatial domain. This analysis is applied to TPIV images from 274°-270° bTDC and consists of 100 cycles/°CA.

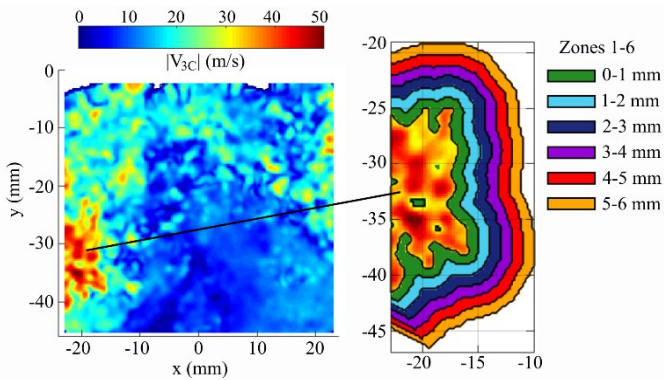

**Figure 15:** Example of spatial domains (1 mm increment regions) extending from the spray-induced jet from which ||S|| and ||Ω|| distributions are extracted.

Figure 16 displays PDFs of $\|S\|$ and $\|\Omega\|$ extracted from within the SIJ and the surrounding expansion zones. Selected °CAs are presented to describe the time progression of the spray-induced turbulence. The PDFs show distinct differences in $\|S\|$ and $\|\Omega\|$ distributions extracted from different zones. Zone 0-1 PDFs consist of values representing the highest $\|S\|$ and $\|\Omega\|$ magnitudes within each spatial domain, including the SIJ domain. This indicates that zone 0-1 contains some of the highest spray-induced turbulence levels observed within the imaging volume. The high $\|S\|$ and $\|\Omega\|$ magnitudes (not only exclusive to zone 0-1) correlate with the accentuated shear layers and strong vorticity pockets that are identified along the immediate SIJ boarder as shown in Figs. 13 and 14. The PDFs for zone 1-2 and the SIJ domain are quite similar at each °CA and represent distributions with the second highest $\|S\|$ and $\|\Omega\|$ magnitudes. As distance increases further from the SIJ, PDFs shift towards lower magnitudes, indicating a decrease in turbulence levels with increasing distance from the SIJ.

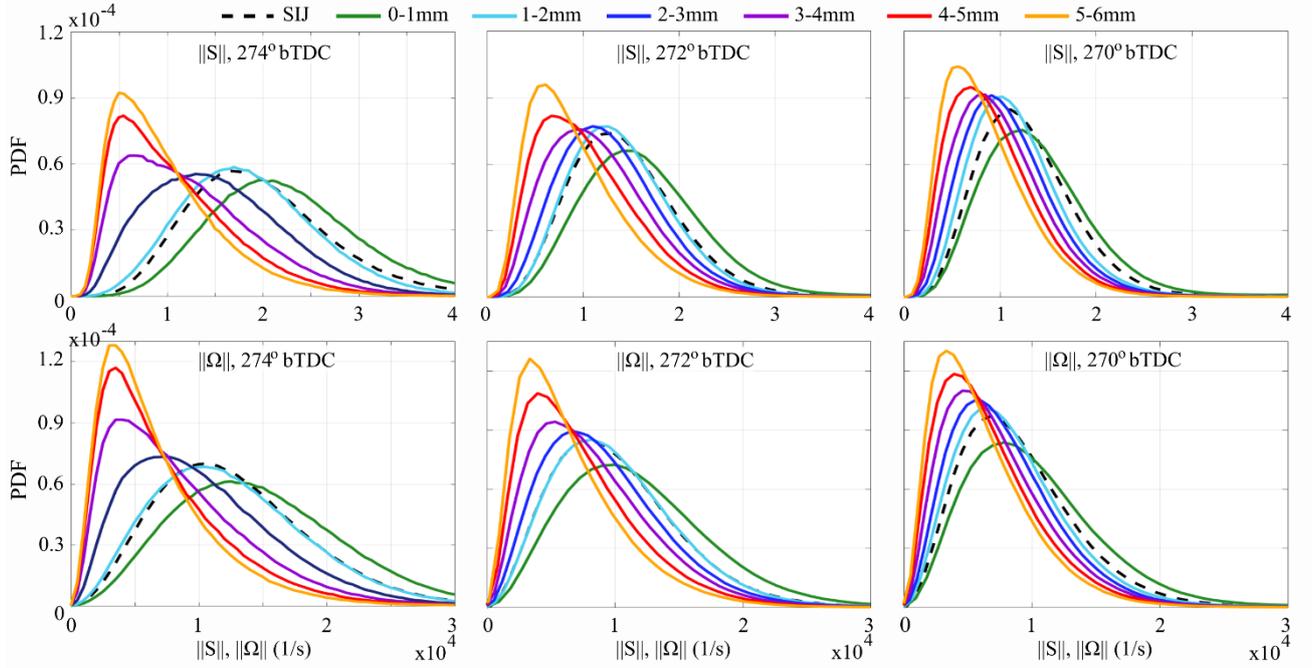

**Figure 16:** PDFs of $\|S\|$ and $\|\Omega\|$ extracted within the SIJ and zones 1-6 extending beyond the SIJ boundary. Statistics are based on 100 cycles and include z-planes from $-2 \leq z \leq 2$ mm.

As °CA progresses, PDFs exhibit narrower distributions that are shifted towards lower magnitudes. This progression towards lower magnitudes represents the decay of spray-induced turbulence that transpires due to molecular diffusion and dissipation (Banerjee and Rutland 2015, Peterson et al. 2017). PDFs at 274° bTDC contain $\|S\|$ and $\|\Omega\|$ that represent the highest turbulence levels for the image timing employed. At each °CA shown in Fig. 16, zone 0-1 continues to exhibit the highest $\|S\|$ and $\|\Omega\|$ values and these values show a monotonic decrease with distance from the SIJ boarder. As °CA progresses, however, the difference between each zone's PDF becomes less pronounced and distributions exhibit more overlap. While the decay of the spray-induced turbulence in time may contribute to the overlap of each zone's PDF, this trend also describes the propensity of the spray-induced turbulence to dissipate spatially as time elapses. The latter occurs as more of the SIJ's kinetic energy is inherently dissipated to the surrounding flow field as time progresses after injection.

The aforementioned behavior of spray-induced turbulence is observed for a single injection event consisting of a small amount of liquid fuel (3.6 mg/cycle, 500 μs injection duration), mimicking a single injection utilized amongst a multi-injection strategy. It is anticipated that spray-induced turbulence levels will be even greater and impact a larger portion of the surrounding flow field for larger amounts of fuel injected and for multiple injection operation.

## 4.4 Relationships between strain rate and vorticity components

Like many jet-like flows, the SIJ imparts turbulence onto the surrounding flow in the form of strong shear layers, which form spatially coherent and temporally evolving vortical flow structures. For fuel sprays, high strain rates and vortical flows play a substantial role in fuel mixing. This section breaks down strain rate and vorticity tensors into their individual components to describe how local strain rates, $S_{ij}$, correlate with nearby vorticity components, $\Omega_k$.

Individual strain rate and vorticity fields are first presented for an individual cycle that contains several isolated vortical structures in the vicinity of the SIJ boundary. This cycle, imaged at 273° bTDC, was chosen because it describes the complex 3D nature of the spray-induced flow and elucidates statistical findings (see Sect. 4.4.2). The flow field for this cycle is shown in Fig. 17; the 3D3C velocity field is imaged on selected planes within the study volume to visualize the SIJ and surrounding regions of strong vorticity. 3D isosurfaces highlight the regions of strong vorticity with $\|\Omega\| \geq 24{,}000$ s$^{-1}$. Detail A-A provides a zoomed view of several vortical flows near the SIJ boundary. In particular, three iso-surfaces are highlighted: $\Omega_A$, $\Omega_B$, and $\Omega_C$.

Evaluation of the flows surrounding $\Omega_A$, $\Omega_B$, and $\Omega_C$ is presented in Figs. 18 and 19. These figures show velocity, vorticity and strain distributions on the 2D planes highlighted in Fig. 17. Normal strain rate distributions are shown in Fig. 18, while shear strain rate distributions are shown in Fig. 19. The selected planes are chosen to discuss unique flow features contributing to the $\Omega_A$, $\Omega_B$, and $\Omega_C$ flow structures. 2D velocity magnitude ($|V_{2C}|$) is presented in the top portion of Fig. 18 to describe the planar velocity distribution in relation to each flow structure. The SIJ boundary and the zone 2 boundary (i.e. 2 mm distance from SIJ) are highlighted in each plane. Vortical flow structures associated with $\Omega_A$, $\Omega_B$ and $\Omega_C$ are overlaid onto each plane and identified by $X_{ij}$, where $X$ refers to the vortical structure and $ij$ refers to the plane. 2D vortex center locations are shown in each plane. Dotted lines, overlaid onto the XY plane, reveal the location of the YZ and XZ planes.

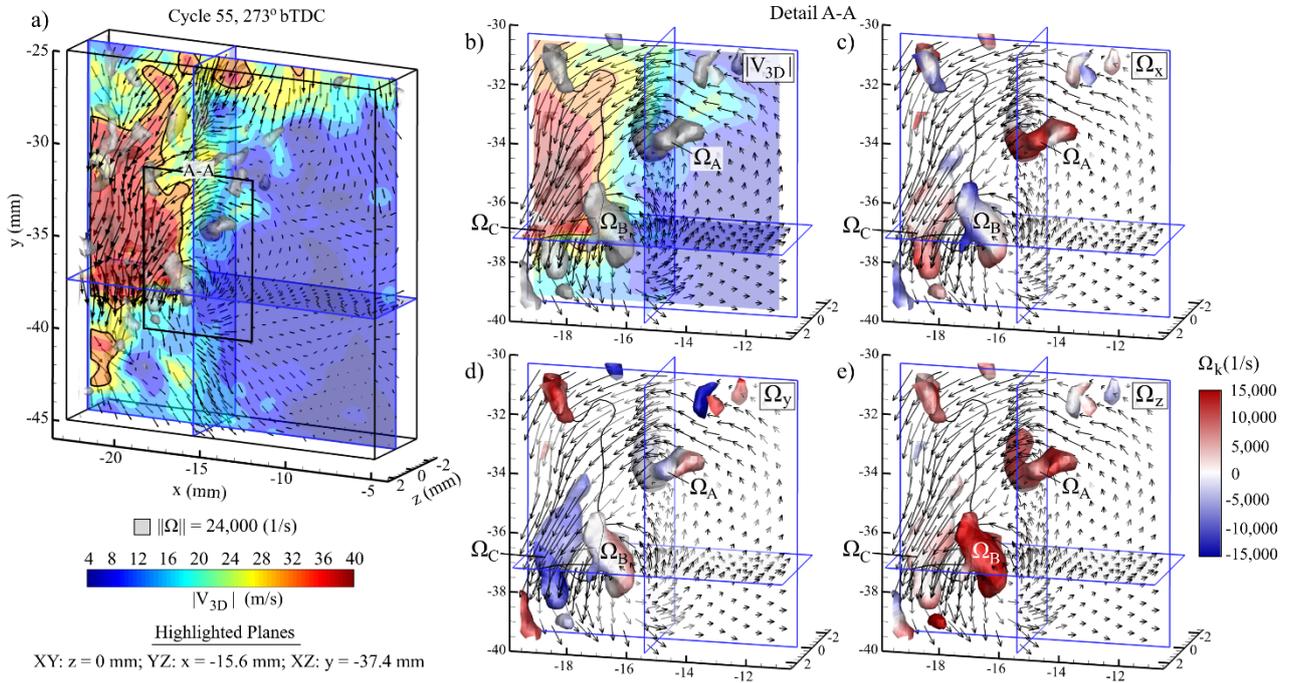

**Figure 17:** 3D3C flow field on selected XY, XZ, and YZ planes for an individual cycle at 273° bTDC. (a) SIJ study volume, (b-e) detail A-A. 3D isosurfaces ($\|\Omega\| = 24{,}000$ 1/s) highlight strong vortical structures ($\Omega_A$, $\Omega_B$ and $\Omega_C$). The 3D iso-surfaces in (c-e) are coloured to show the magnitude of each vorticity vector within the iso-surface. Every 6$^{th}$ [4$^{th}$] vector shown in (a) [(b-e)].

*4.4.1 Vorticity distributions*

$\Omega_A$ consists of a high vorticity region near the edge of zone 2 and is depicted as an individual 3D "Z-shaped" iso-surface. Figure 17c-e shows that $\Omega_A$ is comprised of high (positive) magnitudes of $\Omega_z$ and $\Omega_x$. Figure 18 presents the flows surrounding $\Omega_A$ on the XY and YZ planes. Figure 18 illustrates that $\Omega_A$ is comprised of two distinguishable vortical flow motions, denoted as $A_{xy}$ and $A_{yz}$, which rotate around the z- and x-axes, respectively. $A_{xy}$ and $A_{yz}$ are consequently characterized by regions of high $\Omega_z$ and $\Omega_x$, respectively, and are the dominant vorticity components of $\Omega_A$. Figures 17 and 18 show significant spatial overlap between the $A_{xy}$ and $A_{yz}$ vortical flows as indicated by overlapping regions of high $\Omega_z$ and $\Omega_x$. $A_{yz}$ consists of a small region of elevated $\Omega_x$ in the center of the "Z-shaped" iso-surface, while $A_{xy}$ consists of a larger vortical flow volume extending through the -2 mm $\leq z \leq$ 2 mm domain.

$\Omega_B$ and $\Omega_C$ are two individual vortical structures in close proximity to each other. As shown in Figs. 17c-e, $\Omega_z$ and $\Omega_y$ are the dominant vorticity components of $\Omega_B$ and $\Omega_C$, respectively. The flows surrounding $\Omega_{B/C}$ are analyzed in the XY and XZ planes of Fig. 18. The XY plane shows a strong counter-clockwise vortical flow, denoted as $B_{xy}$, producing the strong (positive) $\Omega_z$ associated with $\Omega_B$. The XZ plane shows a strong clockwise rotating vortex, denoted as $C_{xz}$, producing strong (negative) $\Omega_y$ associated with $\Omega_C$. While $\Omega_B$ and $\Omega_C$ are part of two distinct vortical structures, their iso-surfaces in Fig. 17 are as close as 0.7 mm from each other.

The vorticity and flow fields associated with $\Omega_A$, $\Omega_B$ and $\Omega_C$ emphasize the complex 3D nature of the spray-induced flow, which consists of overlapping or adjacent vortical flows aligned orthogonally from one another.

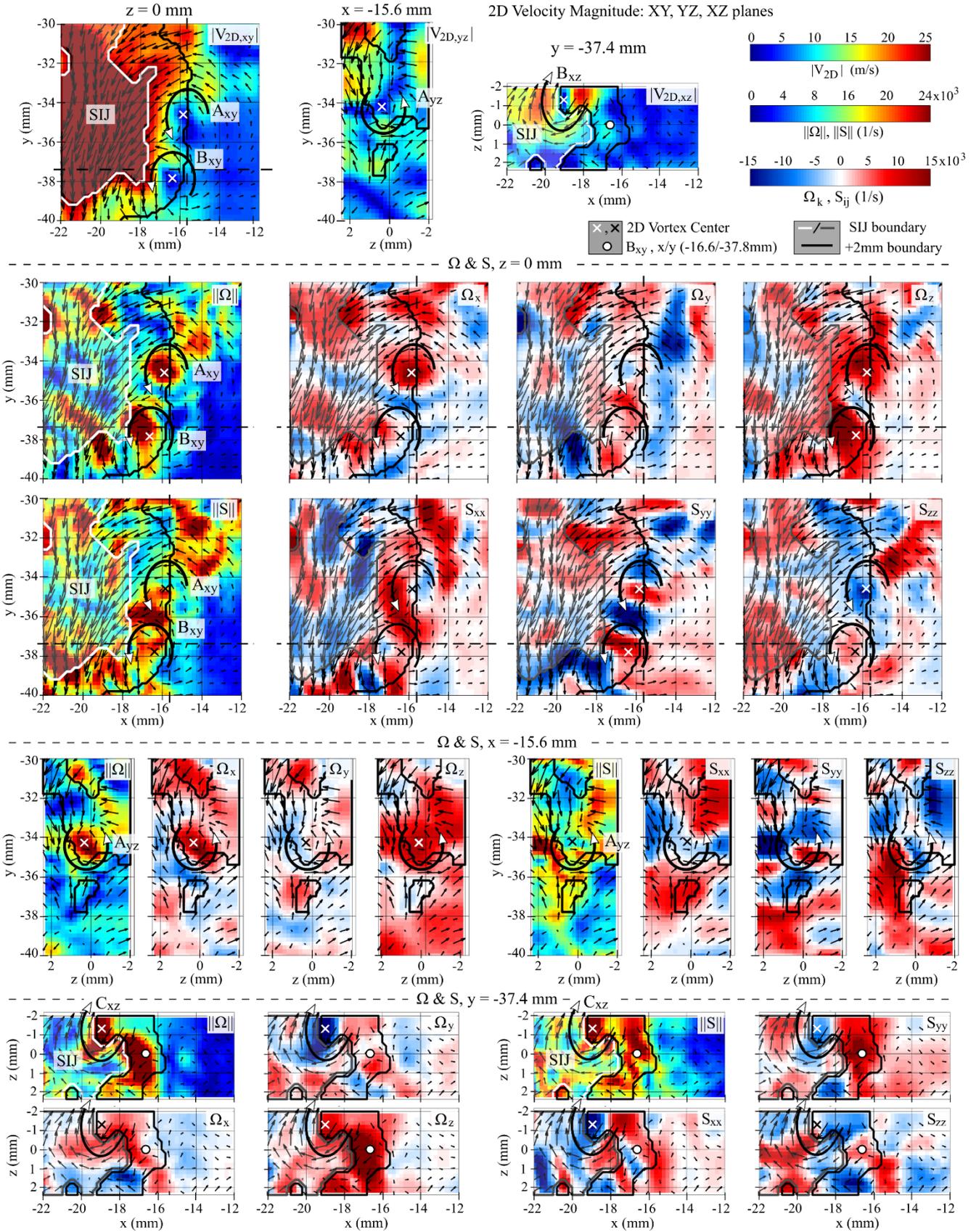

**Figure 18:** 2D velocity magnitude, vorticity, and strain rate components evaluated on selected XYZ planes to describe the spatial distributions surrounding vortical flow structures $A_{xy}$, $A_{yz}$, $B_{xy}$ and $C_{xz}$. The flow corresponds to the individual cycle shown in Fig. 17.

*4.4.2 Strain rate distributions and isosurface density ($\rho_S$)*

Figure 20 is used in conjunction with Figs. 18 and 19 to help quantify the strain rate distributions surrounding the $A_{xy}$, $A_{yz}$, $B_{xy}$ and $C_{xz}$ vortices. $S_{ij}$ magnitudes are extracted from defined regions surrounding individual vortices identified on each 2D plane. An extraction region is defined as a 2D circular region surrounding the vortical structure. The radius of the circular regions (*R*) is defined as:

$$R = R_{avg} + \delta \qquad (19)$$

where $R_{avg}$ is the average radius from the vortex center to the vortex boundary and $\delta$ is an arbitrary distance extending beyond the vortex boundary. For each vortical flow in each *ij* plane, the average $\Omega_k$ magnitude ($|\Omega_{k,avg}|$) was calculated for regions with $|\Omega_k| > 10,000$ s$^{-1}$. The vortex boundary was determined to be the location where $|\Omega_k| = 0.6*|\Omega_{k,avg}|$. This arbitrary boundary accounted for different vorticity strengths and provided a reasonable spatial domain representing elevated $\Omega_k$ magnitudes associated with each vortex. $\delta$ was set to 1 mm to provide a reasonable area extending beyond the vortex boundary that included the periphery flow field. The inserts in Fig. 20 show example extraction regions for the $A_{xy}$, $A_{yz}$, $B_{xy}$ and $C_{xz}$ vortices. While inserts are shown on single planes, this method was applied for each XY, YZ, and XZ plane within the 3D study volume. The plots in Fig. 20 show the strain rate isosurface density ($\rho_S$) within each *R* region. The isosurface density is defined as the area percentage for which the flow exceeds a given value (Zentgraf et al. 2016, Peterson et al. 2017). The $\rho_S$ plots quantitatively describe the highest strain rate magnitudes ($|S_{ij}|$) surrounding each vortical flow.

Strain rate distributions surrounding $\Omega_A$ are discussed first. The XY plane in Fig. 18 shows that strain rate is highest along the periphery of $A_{xy}$, particularly the region between $A_{xy}$ and $B_{xy}$. In this location, the fluid motion from the two co-flowing vortices moves in opposite directions resulting in high strain rates. Individual strain rate distributions reveal that these strain rates are primarily comprised of high (negative) $S_{yy}$ and high (positive) $S_{xx}$. Figure 20 further shows that these in-plane normal strain rate components, $S_{yy}$ and $S_{xx}$, are the strongest strain rates surrounding $A_{xy}$. $S_{xy}$ and $S_{zz}$ are the next highest strain rate components surrounding $A_{xy}$; $S_{xy}$ is comprised of a large (positive) region to the left of $A_{xy}$ (Fig. 19), while (positive) $S_{zz}$ values are highest to the upper right of $A_{xy}$ and (negative) near the vortex center (Fig. 18). $S_{xz}$ and $S_{yz}$ represent the lowest strain rate magnitudes surrounding $A_{xy}$.

For $A_{yz}$, $\|S\|$ values are highest below and to the left of $A_{yz}$. Figures 18 and 20 reveal that normal strain rates ($S_{ii}$) are the most prevalent strain rates surrounding $A_{yz}$. Of these, the in-plane normal strain rates ($S_{yy}$ and $S_{zz}$) are the most prominent, with $S_{yy}$ values being the largest. Of the shear strain rates, the in-plane shear ($S_{yz}$) is the largest, while the remaining shear components are significant lower.

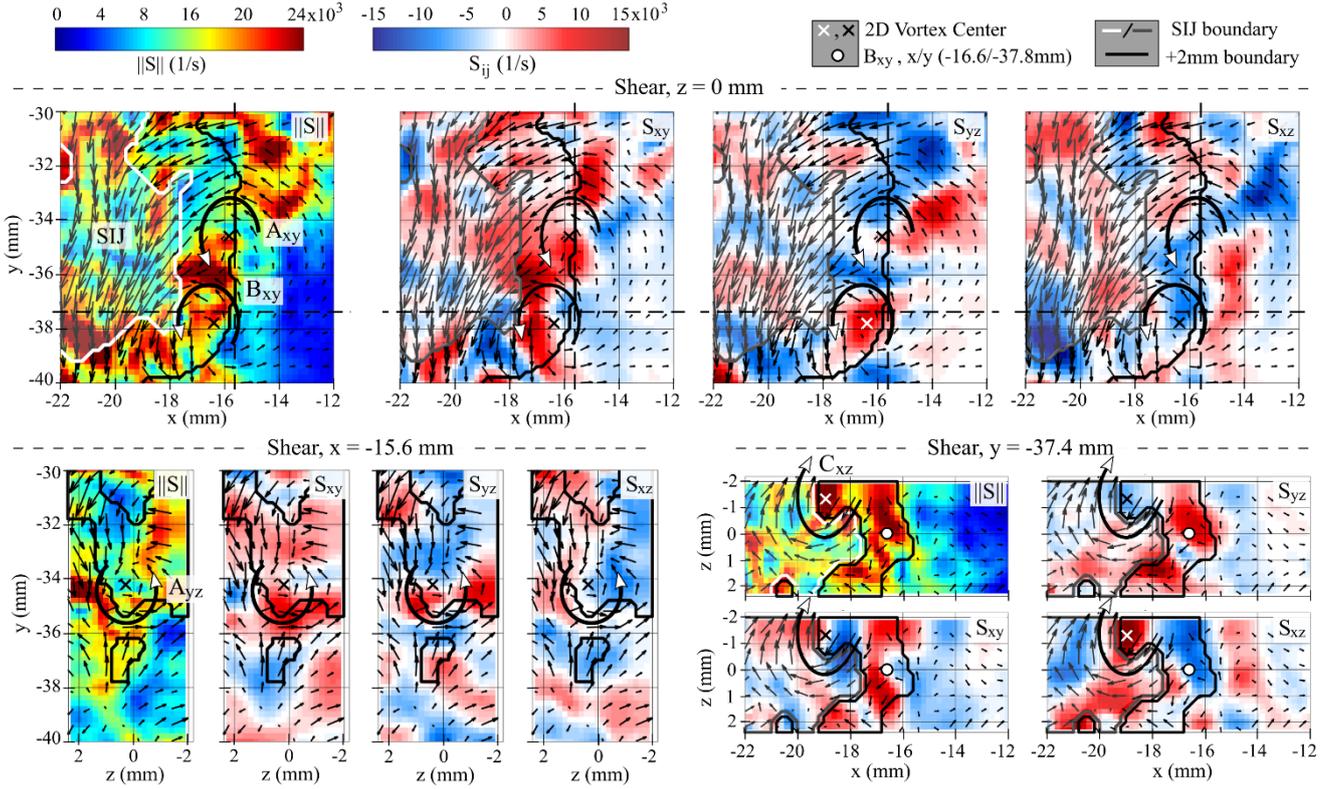

**Figure 19:** Shear strain rate distributions for the planes shown in Figs. 17 and 18.

Evaluating the components of $\Omega_B$ and $\Omega_C$, Fig. 20 reveals that in-plane strain rates surrounding $B_{xy}$ (i.e. $S_{yy}$, $S_{xx}$ and $S_{xy}$) are more predominant than z-related strain rates. $S_{yy}$ is the largest amongst the strain rates surrounding $B_{xy}$; Fig. 18 further shows that regions of highest pos./neg. $S_{yy}$ correspond to regions of highest $\|S\|$. $\rho_S$ plots in Fig. 20 show that $S_{xx}$ and $S_{xy}$ are of similar magnitude and are the second highest strain rates surrounding $B_{xy}$.

For regions surrounding $C_{xz}$, Fig. 20 reveals that normal strain rate components are more prevalent than shear strain rates. Regions of alternating pos./neg. $S_{xx}$ at the $C_{xz}$ center, and alternating pos./neg. $S_{zz}$, along the periphery of $C_{xz}$, contribute to the highest $\|S\|$ surrounding $C_{xz}$. $S_{zz}$ and $S_{xx}$ magnitudes are consequently the highest amongst all $\rho_S$. Regions of alternating high pos./neg. $S_{yy}$ surrounding the $C_{xz}$ center are also substantial, such that $S_{yy}$ magnitudes are the next highest strain rate component. Although shear strain rate distributions are significantly less than normal strain rates, the in-plane shear rate ($S_{xz}$) is amongst the highest of the shear components.

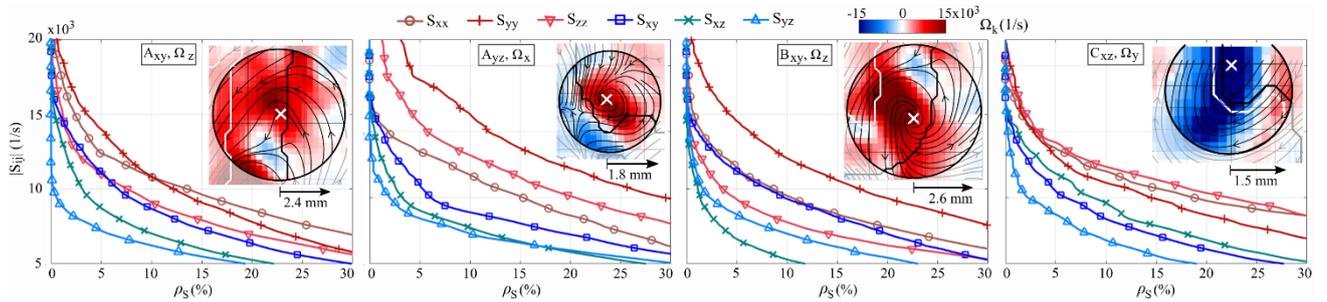

**Figure 20:** Isosurface densities ($\rho_S$) for each $|S_{ij}|$ within the surrounding regions of $A_{xy}$, $A_{yz}$, $B_{xy}$ and $C_{xz}$ vortices. $\rho_S$ values show the percentage of area exceeding $|S_{ij}|$.

There are several trends observed from the data presented in Fig. 20. Within each *ij* plane, the in-plane normal strain rates ($S_{ii}$, $S_{jj}$) are consistently dominant surrounding $\Omega_k$ vortical structures. For the planes including the vertical direction, $S_{yy}$ typically has the largest magnitude. Hence, in the immediate regions surrounding the SIJ, gradients of y-velocity (i.e. the most dominant SIJ velocity component) are largest in the direction of jet travel. Of the shear strain rates, the in-plane shear ($S_{ij}$) is often the most dominant shear component surrounding $\Omega_k$. 3D effects are also important; the out-of-plane normal strain rate ($S_{kk}$) is often greater than the in-plane shear strain rate. Overlapping regions of vortices rotating around different axes further emphasizes the strong three-dimensionality of the flow. Such overlapping vortices also cause variations in the aforementioned trends, which makes direct correlations difficult. For example, regions of high $S_{yy}$ associated with $B_{xy}$ also surround $C_{xz}$ (see XY and XZ planes, Fig. 18) such that $S_{yy}$ magnitudes for $C_{xz}$ in Fig. 20 are nearly similar to $S_{xx}$ and $S_{zz}$. Similar arguments can be made for regions of high $S_{xx}$ (and moderate $S_{xy}$) associated with $A_{xy}$, which overlap with surrounding regions of $A_{yz}$.

Figure 21 shows $\rho_S$ distributions surrounding $\Omega_x$, $\Omega_y$, and $\Omega_z$ vortices identified in all cycles for selected °CAs. Similar trends exist as those presented in Fig. 20 for individual vortex structures. Namely:

- $S_{ii}$ and $S_{jj}$ strain rates are the most dominant strain components surrounding $\Omega_k$ vortical structures.
- For $\Omega_k$ components including the vertical velocity, $S_{yy}$ is consistently the dominating strain rate component.
- Out-of-plane normal strain rates ($S_{kk}$) are typically greater than in-plane shear strain rate ($S_{ij}$). This emphasizes the importance of 3D effects, which may include overlapping vortices rotating around different axes.
- Amongst the shear strain rate components, $S_{ij}$ is often the highest in magnitude surrounding $\Omega_k$.

As °CA progresses, strain rate components gradually decrease. This decrease represents the decay of spray-induced turbulence with time due to molecular diffusion and dissipation (Bharadwaj et al. 2009, Peterson et al. 2017). The aforementioned trends remain consistent as strain rates decay with °CA. As strain rates decay with °CA, many of the $\rho_S$ curves begin to exhibit more overlap; by 270° bTDC, the out-of-plane normal strain rate magnitudes vastly approach those of in-plane shear strain rate magnitudes.

Amongst the trends exhibited, $S_{yy}$ remains the dominant strain rate surrounding $\Omega_{z/x}$ within vertical planes. The downward motion of the SIJ and downward motion of the piston (i.e. volume expanding in y-direction) likely contribute to the high $S_{yy}$ magnitudes observed in the imaging volume. As these predominant downward motions exist, it is anticipated that a $\Omega_z$ or $\Omega_x$ vortex will become stretched in the y-direction (Kerswell 2002). This would lead to high $S_{yy}$ strain rates along the periphery of vortices as evidenced in Fig. 18. Another mechanism contributing to high $S_{yy}$, is the impingement of the SIJ on lower velocity regions, which creates large $\partial v/\partial y$ gradients along the lower boarder of the SIJ (e.g. southwest of $B_{xy}$, Fig. 18).

It is important to emphasize that the analyses presented in this section have limitations. The z-volume is limited (4 mm) and observed trends may be biased as more x-y- information is available than z-information. These trends may also be specific to the imaging volume centered around the cylinder axis (z = 0 mm). Findings may differ in other cylinder locations. Nonetheless, the presented measurements enable us to quantify and understand the spray-induced turbulence within the FOV. Furthermore, the measurements are intended to provide valuable data for numerical modelling. Once validated, simulations can be used to understand turbulence in other locations and their implications on turbulent mixing.

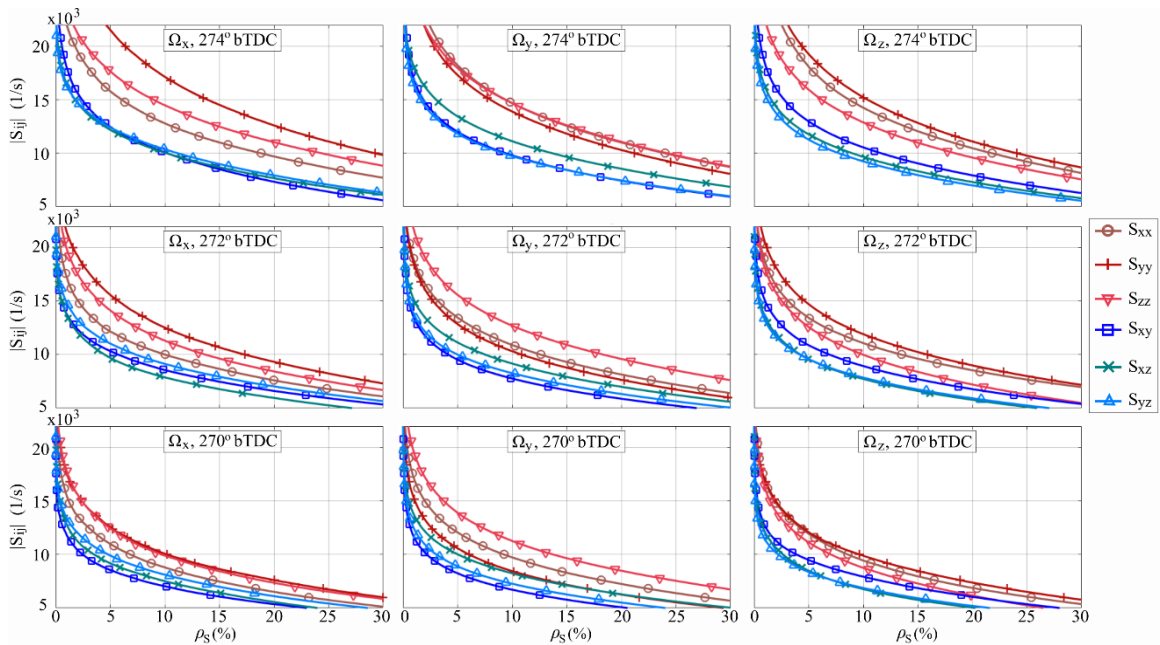

**Figure 21:** Isosurface densities ($\rho_S$) of $|S_{ij}|$ surrounding $\Omega_x$, $\Omega_y$, and $\Omega_z$ based on 100 cycles.

## 5 Conclusions

Tomographic PIV measurements were performed to study the spray-induced turbulence during the intake stroke of a spray-guided DISI optical engine. Fuel injection from a hollow cone injector produced a high-velocity jet in the far field of the central symmetry plane. This spray-induced jet, comprised of air and droplets, travels downwards through the cylinder and imparts turbulence onto the surrounding in-cylinder flow. TPIV measurements were performed after a single-fuel injection when particle densities were suited for accurate 3D particle reconstruction.

HS-PIV measurements at 4.8 kHz were combined with TPIV at 3.3 Hz to provide a time-history of the 2D2C flow field preceding TPIV images. Mie scattering images were evaluated to describe droplet distributions suitable for TPIV particle reconstruction. A simplified methodology utilizing geometrical optic arguments was applied to estimate particle diameter and response times to reveal that particles are expected to follow the gas flow. HS-PIV was further used to validate TPIV in the central symmetry plane. A comprehensive uncertainty analysis was performed for TPIV to assess the uncertainty associated with individual vorticity and strain rate components.

TPIV measurements were first evaluated to assess statistical moments of the 3D flow, which revealed that the highest turbulence levels were associated with the high-velocity SIJ. A velocity threshold method was used to identify the spatial domain of the SIJ within individual TPIV images. Spray-induced turbulence was analyzed by evaluating $\|S\|$ and $\|\Omega\|$ fields in relation to the SIJ domain. TPIV images revealed the presence of strong shear layers (visualized by high $\|S\|$) and pockets of elevated vorticity along the immediate boundary of the SIJ. $\|S\|$ and $\|\Omega\|$ magnitudes were extracted from spatial domains extending in 1 mm increments from the SIJ to quantify the spatial domain of the spray-induced turbulence. Zone 0-1mm contained the highest spray-induced turbulence levels observed within the imaging volume. Turbulence levels decrease with radial distance from the SIJ boarder. As °CA progresses, PDFs described the propensity of the spray-induced turbulence to dissipate spatially from the SIJ.

Individual strain rate and vorticity components were evaluated to describe how local strain rate components, $S_{ij}$, correlate with nearby vorticity components, $\Omega_k$. Analysis reveals the complex 3D flow geometries with strong $\|\Omega\|$ flow structures comprising of many overlapping vorticities rotating around separate axes. Accordingly, $\Omega_k$ flow structures are not characterized by a single $S_{ij}$ component, and the order of which $|S_{ij}|$ components are largest varies for individual $\Omega_k$ flows. Overall, normal strain rate components, $S_{ii}$, were often the largest surrounding $\Omega_k$ flows, with in-plane normal components being the most dominant. $S_{yy}$ was consistently the largest for $\Omega_k$ flows involving the vertical direction. The strong vertical motion of the SIJ likely contributes to the high $S_{yy}$ values observed in the images. 3D effects are important as out-of-plane normal strain rates, $S_{kk}$, were typically greater than in-plane shear strain rates, $S_{ij}$. Of the shear strain rates, in-plane shear was consistently higher than out-of-plane shear components.

While findings are limited to small z-volumes and injection is comprised of a single-injection with a small amount of fuel (3.6 mg/cycle), the unique ability to analyse the 3D spray-induced flow is conducive to understand flows responsible for local fuel air mixing. This work is also intended to provide valuable data and comparison metrics for numerical modelling development.

## 6. Acknowledgments

B. Peterson kindly acknowledges financial support from the European Research Council (grant #759546) and EPSRC (EP/P020593/1). B. Böhm kindly acknowledges financial support from the Deutsche Forschungsgemeinschaft (SFB/Transregio 150). A. Dreizler kindly acknowledges financial support from the Gottfried Wilhelm Leibniz program. The authors are also especially grateful to LaVision for borrowing of equipment for the experimental campaign. The authors are also thankful to Dirk Michaelis from LaVision for useful discussions about suitable particle density distributions for accurate tomographic reconstruction.